\documentclass[12pt]{article}

\usepackage[utf8]{inputenc}
\usepackage{comment}
\usepackage{times}
\usepackage{graphics,graphicx}
\usepackage{color}
\usepackage{xcolor,soul}
\usepackage{amsmath,amsthm,amssymb}
\usepackage{mathtools}
\usepackage{stmaryrd}
\usepackage[normalem]{ulem}
\usepackage{hyperref}

\usepackage[font=small,labelfont=bf]{caption}

\usepackage{tikz-cd}

\definecolor{ao(english)}{rgb}{0.0, 0.5, 0.0}

\newenvironment{svmultproof}{\begin{proof}}{\end{proof}}

\newtheorem{theorem}{Theorem}

\begin{document}
\title{Network Bypasses Sustain Complexity}

\author{
Ernesto Estrada$^{1}$\footnote{estrada@ifisc.uib-csic.es}, Jes\'us G\'omez-Garde\~nes$^{2,3}$\footnote{gardenes@gmail.com}, Lucas Lacasa$^{1}$\footnote{lucas@ifisc.uib-csic.es}\\
\normalsize{$^{1}$Institute for Cross-Disciplinary Physics and Complex Systems (IFISC)}\\
\normalsize{CSIC-UIB, Palma de Mallorca, Spain}\\
\normalsize{$^{2}$Department of Condensed Matter Physics, University of Zaragoza, E-50009 Zaragoza, Spain}\\
\normalsize{$^{3}$GOTHAM Lab -- Institute for Biocomputation and Physics of Complex Systems (BIFI)}\\
 \normalsize{University of Zaragoza, E-50018 Zaragoza, Spain}}

\baselineskip24pt

\maketitle


\begin{abstract}
Real-world networks are neither regular nor random, a fact elegantly explained by mechanisms such as the Watts-Strogatz or the Barab\'asi-Albert models, among others. Both mechanisms naturally create shortcuts and hubs, {which while enhancing network's connectivity, also might yield several undesired navigational effects: they tend to be overused during geodesic navigational processes --making the networks fragile-- and provide suboptimal routes for diffusive-like navigation.}  Why, then, networks with complex topologies are ubiquitous? Here we unveil that these models also entropically generate network bypasses: alternative routes to shortest paths which are topologically longer but easier to navigate. We develop a mathematical theory that elucidates the emergence and consolidation of network bypasses and measure their navigability gain. We apply our theory to a wide range of real-world networks and find that they sustain complexity by different amounts of network bypasses. At the top of this complexity ranking we found the human brain, which points out the importance of these results to understand the plasticity of complex systems.
\end{abstract}

\newpage

\section{Introduction}

The advent of Network Science \cite{book1, book2} was marked by the urgent need to decipher simple and local mechanistic models underlying the self-organized formation and growth of natural and artificial real-world networks, models  
able to parsimoniously account for large-scale structural patterns systematically deviating from stylized ones such as purely ordered lattices or purely random graphs. 
Two such celebrated  models, aiming to explain the ubiquitous real-world patterns of  ``small-worldness'' (SW) and ``scale-freeness'' were  proposed in seminal contributions by Watts and Strogatz \cite{WS} and Barab\'asi and Albert (BA) \cite{BA}, respectively. 
The resulting network topologies of SW and BA networks  --poised between order and disorder at the statistical level-- were coined as \textquoteleft complex\textquoteright{}. Here we give a special attention to these as they are paradigmatic mechanisms that create complexity through heterogeneization,
although we acknowledge that other patterns \cite{book2} --e.g. communities, assortative and dissassortative mixing, triadic closure, etc-- are also relevant in this context.\\
 
\noindent Indeed, what is complex in a complex network? Conceptually, system complexification \cite{tranquillo} may occur via different types of mechanisms including symbiosis, exaptation or structural deepening to cite some. The latter concept of structural deepening \cite{arthur}, which we adopt here, focuses on the situation where the efficiency of an existing function in the system is increased as the system complexifies, where a higher efficiency is usually interpreted in terms of performing the same function using less available energy. Accordingly, a network with a structure poised between total order (lattice) and pure disorder (random graph), such as SW and BA networks as well as many networks in the real-world, is compatible with the existence of a structural deepening mechanism which improves the communication efficiency between the nodes in the networks. Identifying a quantitative proxy that characterizes such structural deepening mechanism in networks remains, however, an open problem, and constitutes the first motivation of this work. As a matter of fact, the Watts-Strogatz mechanism does not provide a clear-cut definition of what a SW network is --only a certain range of network\textquoteright s mean path length and clustering coefficient--, indicating that neither of these two network properties are quantitative proxies of a potential structural deepening mechanism. Similarly, the extensive zoology of degree distributions existing in empirical networks \cite{controversy1,controversy2} points to the fact that observing scale-freeness is not in itself enough to indicate the existence or not of a structural deepening mechanism. Other network properties, such as the node-based fractal dimension (NFD),  the node-based multifractal analysis (NMFA), the structural distance, or the degree of complexity \cite{xiao,xue} suffer from similar problems, and e.g. fail to identify a specific point within the SW region where structural deepening is maximized.\\

\noindent And yet, networks serve the purpose of facilitating the communication between otherwise isolated entities of a complex system. Therefore, if a structural deepening mechanism exists in the evolution of a network it is likely that it involves an improvement of some communication efficiency. SW and BA-type mechanisms indeed tend to generate networks with enhanced connectivity \cite{xiao} (a form of structural deepening) which are robust against random failures \cite{tolerance, xue}, what in principle could explain the ubiquity of these mechanisms and the resulting macroscopic patterns, even if quantifying such complexity has proven elusive.\\ 

\noindent However, observe that the SW mechanism reduces mean path length simply by creating path shortcuts, making enhanced connectivity overly dependent --and thus, fragile-- on them. Likewise, in BA-like networks, shortest paths often involve hubs, and these networks are known to be extremely fragile against failure of hubs \cite{porter} or jamming  \cite{jamming, jamming2, jamming3, jamming4},  potentially inducing a failure cascade which can severely harm the macroscopic network's function.\\

\noindent Why, then, complex networks are ubiquitously observed?
First, note that walkers navigating a network do not necessarily have full information of the network structure, and geodesic navigation is indeed a global optimization problem \cite{go} that, accordingly, ``blind'' walkers cannot perform. Second, such blind walkers typically undergo diffusion-like navigation and such parsimonious navigation strategy can lead walkers to `diffuse out' and get lost easily if attempting to follow shortest paths, as these tend to have higher degree nodes\footnote{A node of degree $k$ potentially connects $k(k-1)/2$ pairs of nodes by shortest paths of length two. Longer SP also use them to connect other pairs of nodes. Thus, the higher the degree of a node the higher the number of SP crossing that node.}. Accordingly non-geodesic navigational strategies have been proposed \cite{r5,r6,r7,r8,r9, r10, r11}, usually providing heuristic recipes based on local network information available (such as the degree \cite{r5,r6,r7} or the matching index \cite{r8}). Solving the apparent dilemma between the prevalence of complex network architectures --underpinned by WS and BA mechanisms among others-- with structural deepening related to enhanced communication capacity requires to find parsimonious mechanisms which can mitigate the undesired effects of geodesic navigability, and this is the second motivation of our work.\\

\noindent Our contention in this work is that as a network complexifies, it is capable to mitigate the impact of the undesired geodesic navigability issues by structural deepening mechanisms which favor the consolidation of network bypasses: alternative routes to mere geodesic navigation that (i) decrease the tendency of \textquoteleft getting lost\textquoteright{} by blind walkers, and (ii) if needed, can also be used by non-blind walkers to avoid problematic links and nodes, therefore allowing the overall connectivity to be maintained and the network to be robust against failure of shortcuts and hubs.\\ 
\newpage

\noindent In what follows we start from first principles and develop a theory to define and detect the emergence of network bypasses in both synthetic and real-world networks, and quantify their associated gain and impact in terms of network navigability. Our theory is based on a network geometrization by which initially unweighted edges and paths acquire an effective weight --an effective length, or cost-- induced solely by the topology of the surrounding network's structure. Network bypasses then emerge as geodesic paths in the geometrized network (i.e. they are the solutions of a new type of topology-induced minimum-cost path optimization problem \cite{flow}), and in many cases we show they don't coincide with the shortest paths of the original network.
We also show that (i) the emergence of these network bypasses is an unavoidable (entropic) byproduct of the WS and BA mechanisms themselves, and that (ii) the effect of these bypasses is optimally emphasized when networks fall in a specific point of SW regime and an intermediate edge density in the sparse regime for BA-like networks, thus finding a quantitative proxy for structural deepening. We also certify that (iii) network bypasses indeed provide source-destination routes with better navigation properties for diffusive-like blind walkers than geodesic routes, and finally rank and discuss the emergence of network bypasses and their associated navigability gain in a range of real-world networks.\\

\section{Results}

\begin{figure}[t!]
\begin{centering}
\includegraphics[width=0.95\textwidth]{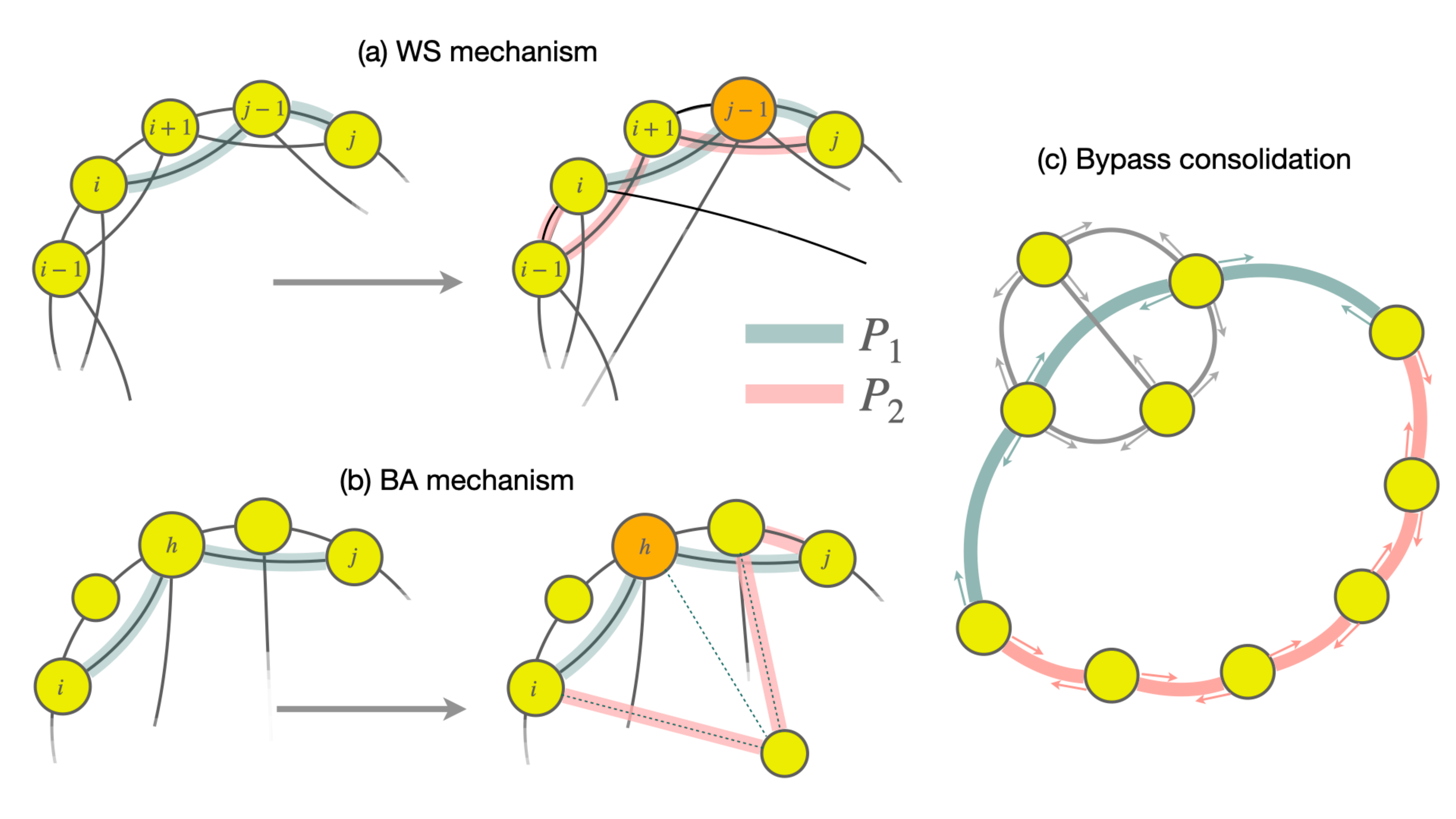}
\par\end{centering}
\caption{(Panel a) Illustration of the effects of the edge rewiring process in the Watts-Strogatz
model on the paths connecting two arbitrary vertices of the resulting
graph: the shortest path $P_1$ (blue) can by bypassed by the path $P_2$ (pink), topologically longer but with a lower energetic cost. (Panel b) Similar phenomenon happens when the node $h$ becomes a hub after a rich-get-richer mechanism. Shortest path $P_1$ (blue) typically cross the hub but, with a sufficiently large mean degree, other paths such as $P_2$ (pink) can bypass the shortest path $P_1$, allowing alternative routes when hubs reach capacity and become saturated or damaged. (Panel c) Navigational dilemma embedded in a network: the blue path $P_1$ is the shortest path, but it turns out that the pink `braquistochronic' path $P_2$ is more advantageous as it avoids congestion and is less resistive (see SI section S5 for an explicit calculation).}
\label{WS_sketch_both}
\end{figure}

To fix the intuition, let us begin by illustrating two situations in simple graphs that highlight the importance of bypasses in the operation of a network that harbors transportation and propagation of signals and information. To this aim, we initially consider a particle hopping between the nodes of a network created via the WS model \cite{WS}, and we focus on the propagation of the particle between nodes $i$ and $j$ (see Panel (a) in Fig. \ref{WS_sketch_both}).  
Starting with rewiring probability $p=0$ we have a circulant graph $\cal G$, and the path $P_{1}=\{ i,j-1,j\}$ of length 2 (highlighted in blue) is a shortest path connecting $i$ and $j$. Mimicking the action of WS-like mechanism kicking in, {the edge $e=(i,i+1)$ of $\cal G$ is randomly rewired, and subsequently another edge is also randomly rewired, so that node $j-1$ now receives an edge from a `distant` node. In the resulting graph ${\cal G}'$, vertex $i+1$  drops}
its degree by one, whereas vertex $j-1$ increases its degree. This situation creates a small degree heterogeneity in the graph ${\cal G}'$ which did not exist in the circulant graph $\cal G$: node $j-1$ now participates in many more shortest paths starting elsewhere and ending at vertex $j-1$. Accordingly, the length-2 path $P_{1}$, in practice, might not be the ``best'' route to connect $i$ and $j$, even if it is still the shortest path, topologically speaking. {For instance, 
a random walker choosing $P_{1}$ has a higher likelihood of ``diffusing out'' through $j-1$, thus hardly reaching the destination. Likewise, geodesic navigation will make $j-1$ systematically overused, leading to a higher chance of damage or jamming.}
In turn, the length-3 path $P_2=\{ i,i-1,i+1,j\} $ (highlighted in pink), while being topologically longer than $P_1$, contains node $i+1$ whose degree is at the same time lower than the average and also avoids $j-1$, hence can be seen as a potentially more ballistic route that avoids a potentially problematic $j-1$ and still connects $i$ and $j$.\\

A similar situation is depicted in Panel (b) of Fig.\ref{WS_sketch_both} where node $h$ becomes a hub via a rich-get-richer (i.e. BA-like) mechanism. The shortest path between $i$ and $j$ (highlighted in blue) will again be more prone for the walker to get lost due to the presence of a high-degree node, 
once the BA mechanism enhances such heterogeneity. Now, if the network supports a sufficiently large\footnote{yet sufficiently small so that the network is in the sparse regime} mean degree --i.e. if the network allows more edges to be formed than a spanning tree--, then other routes can emerge, bypassing the hub (pink path).\\

The two examples illustrated in Panels (a) and (b)  Fig.~\ref{WS_sketch_both} raise the question of whether a particle would ``prefer" to travel from $i$ to $j$ via the {shortest --albeit with higher uncertainty to reach the destination-- path $P_{1}$ or along the slightly longer but more ballistic --smaller uncertainty-- alternative path $P_{2}$.} In Panel (c) of the same figure we illustrate such conundrum, where two alternative routes (a shortest path $P_1$, in blue, and a topologically longer one $P_2$, in pink) are highlighted. 
Intuition tell us that there should be a trade-off: sometimes $P_{1}$ is to be preferred, sometimes $P_{2}$ is a contingently better option. Extending this situation to a network-growth mechanism, this suggests that the creation of shortcuts (SW) and hubs (BA) should be sustained by the emergence of some alternative paths bypassing these, with structural deepening effects that would reach a maximum impact for a specific rewiring probability $p$ (SW) as well as specific hub abundance  (BA). In what follows we introduce a formalism that puts these questions and their general solution in a solid grounding.

\subsection*{The concept of {\em Resistive Paths}}

Starting from first principles, the possible trajectories that a hopping particle can perform over a network ${\cal G}=(V,E)$ of $|V|=n$ nodes
with binary adjacency matrix ${\bf A}=\{A_{ij}\}_{i,j=1}^n$ 
can be enumerated by computing the powers of ${\bf A}$. A natural way to penalize longer trajectories connecting the same initial and end nodes is to properly weight them
\begin{equation}
\textbf{G}(\beta)=e^{\beta{\bf A}};\ G_{ij}(\beta)=\sum_{l=0}^{\infty}\frac{\beta^l \left({\bf A}^l \right)_{ij}}{l!}=\left(e^{\beta{\bf A}}\right)_{ij}\;,
\end{equation}
where $\beta$ is an empirical parameter. This expression is known as the communicability function of a graph \cite{comm, oscillators}. While originally being a purely combinatorial expression that encapsulates the contributions of different walks in a graph, $\textbf{G}(\beta)$ indeed emerges as a central matrix when analysing a wide variety of dynamics on graphs \cite{oscillators,subgraphcentrality,LTE,Bartesaghi,Arola} (see SI S1.1 for details and S1.2 for a derivation of $\textbf{G}(\beta)$ as the actual propagator in a specific case with Hamiltonian dynamics).
Nowadays communicability is applied across a range of disciplines, from neuroscience \cite{neuro0, neuro01, neuro1, neuro2, neuro3, neuro4, neuro5,neuro6} or cancer research \cite{cancer1} to ecology \cite{ecology1} or economics \cite{economics1}, to cite a few.\\

\noindent {While this operator naturally emerges on relation to different types of dynamics on networks, in this work we shall highlight that it is fundamentally a combinatorial one, and is not a priori derived from any concrete dynamics running on the network. In other words, while we will consider that there is some kind of generic propagation --let it be information, electrons, or other types of particles hopping through the network--,  the theory presented hereafter does not require to specify which dynamical equations rule such propagation, as we focus on the structural (topological) constraints which generally affect such propagation. By analogy to the cases discussed in SI S1.1, S1.2 and \cite{oscillators}, we call $G_{ij}$ the structural propagator, which parsimoniously captures the role that the network's architecture plays in $j$ receiving particles sent from $i$}. Similarly, $G_{ii}$ accounts for how much a node $i$ structurally retains an item at it, as the item returns to $i$ infinitely often. For a particle initially located at the node $i$ the difference:
\begin{equation}
R_{i\rightarrow j}(\beta)=G_{ii}(\beta)-G_{ij}(\beta)
\label{resist}
\end{equation}
 {accounts for the opposition offered by the network structure to the directional displacement of a particle} sent by node $i$ to the node $j$, where the smaller the value of $R_{i\rightarrow j}$, the higher
the probability that the particle does not get trapped at the origin $i$ and can propagate to node $j$, {\em i.e.}, there are more conductive walks between $i$ and $j$ than those returning back to the origin. In order to account for the resistance of the displacements between any pair of nodes we should take into account the two possible directions of their mutual communication ($i\rightarrow j$ and $j\rightarrow i$). To this aim, one can symmetrize (\ref{resist}) to define the \textit{communication resistance} between nodes $i$ and $j$ as: $\xi_{ij}(\beta)  \coloneqq (R_{ij}(\beta)+R_{ji}(\beta))^{1/2}$. From the definition of the communicability function, and setting $\beta=1$ without loss of generality, we obtain that the \textit{communication resistance} reads:
\begin{equation}
 \xi_{ij}^2=\sum_{m=1}^{n}e^{\lambda_m}\left(\left(\psi_{m}\right)_i-\left(\psi_{m}\right)_j\right)^2\;.
\end{equation}
where $\left(\psi_{m}\right)_i$ is the $i$-th entry of the eigenvector associated to the $m$-th eigenvalue ($\lambda_m$) of ${\bf A}$.
We rigorously proved that $\xi_{ij}$ is an Euclidean distance (see SI section S2 por a proof). {Conceptually, $\xi_{ij}$ is a measure of the network resistance to a flow between $i$ and $j$. Recently  \cite{estrada_proof}, it was proven that this communicability distance --and every spherical euclidean distance-- is the effective resistance between nodes in a network with given edge weights.}

\subsection*{Network Geometrization and {\em Resistive Shortest Paths}} 

{Since $\xi_{ij}$ is an Euclidean distance and particles motion is confined to the network edges, we can proceed to the geometrization of the network \cite{geom1, geom2}. To this aim we first transform every edge of the graph into a compact 1-dimensional
manifold. That is, for an edge $e=\left\{ i,j\right\} $ we consider
the boundary of the manifold to be $\partial e=i\cup j$. Then,
each edge $e$ inherits a metric $g_{e}$ such that $(e,g_{e})$ is
isometric to a finite interval $[0,L(e)]$ of the real line with the
standard metric, where the length $L(e)$ is given by the
communicability distance of the corresponding edge, i.e. $L(e)\equiv \xi_e=\xi_{ij}$. Finally, the distance
metric on the edges is extended to the full graph via infima of lengths
of curves in the geometrization of $\cal G$, such that the graph becomes a 
metrically complete length space \cite{geom2}.}\\

\noindent Equipped with this geometrization, we can now define two different types of lengths for any given path $\mathfrak{p}(s\to t)=(s,\dots,t)$ connecting nodes
$s$ and $t$ in the network.  First, the \textit{topological length} $\ell_{\mathfrak{p}(s\to t)}$ of this path is just the number of edges
in it. Among all paths $\{\mathfrak{p}(s\to t)\}$ connecting $s$ and $t$, the one with the minimum length is denoted the shortest path $\text{SP}(s,t)$ as 
\begin{equation}
\text{SP}(s,t)=\text{argmin}_{\mathfrak{p}(s\to t)}[\ell_{\mathfrak{p}(s\to t)}]
\label{eq:SP}
\end{equation}
Observe that Eq.\ref{eq:SP} can have more than one solution, specially for large networks (see SI S4).\\
Second, and based on the geometrization induced by the communicability resistance above, we also define an \textit{effective length} $\mathbb{L}_{{\mathfrak{p}(s\to t)}}$ by summing the induced length of each of the links involved in $\mathfrak{p}(s\to t)$:
\begin{equation}
\mathbb{L}_{{\mathfrak{p}(s\to t)}}=\sum_{\left(i,j\right)\in E\in \mathfrak{p}(s\to t)}{\xi_{ij}}\;. 
\label{bigL}
\end{equation}
At odds with $\ell_{\mathfrak{p}(s\to t)}$, which blindly assigns the same length (unity) to every edge of the network, $\mathbb{L}_{{\mathfrak{p}(s\to t)}}$ takes into account the topological neighborhoods of each of the nodes in the path and the associated likelihood that the particle might diffuse out of the path, accordingly. Likewise, it penalises paths for which particles take naturally more time to travel due to the structure of the network in which the path is embedded in. The specific path connecting $s$ and $t$ that minimizes this effective length is denoted the
{\em Shortest Resistive Path}  $\text{SRP}(s,t)$, defined as:
\begin{equation}
\text{SRP}(s,t)=\text{argmin}_{\mathfrak{p}(s\to t)}[\mathbb{L}_{\mathfrak{p}(s\to t)}].
\label{eq:SRP}
\end{equation}
We are now ready to quantify (i) the emergence of potential bypasses --i.e. the proliferation of non-SP between any two nodes-- and (ii) decide in a principled way when this path redundancy become relevant to the network function --something that, we advance, will happen when SRPs start to differ from SPs--.

\begin{figure}[t!]
\begin{centering}
\includegraphics[width=0.9\textwidth]{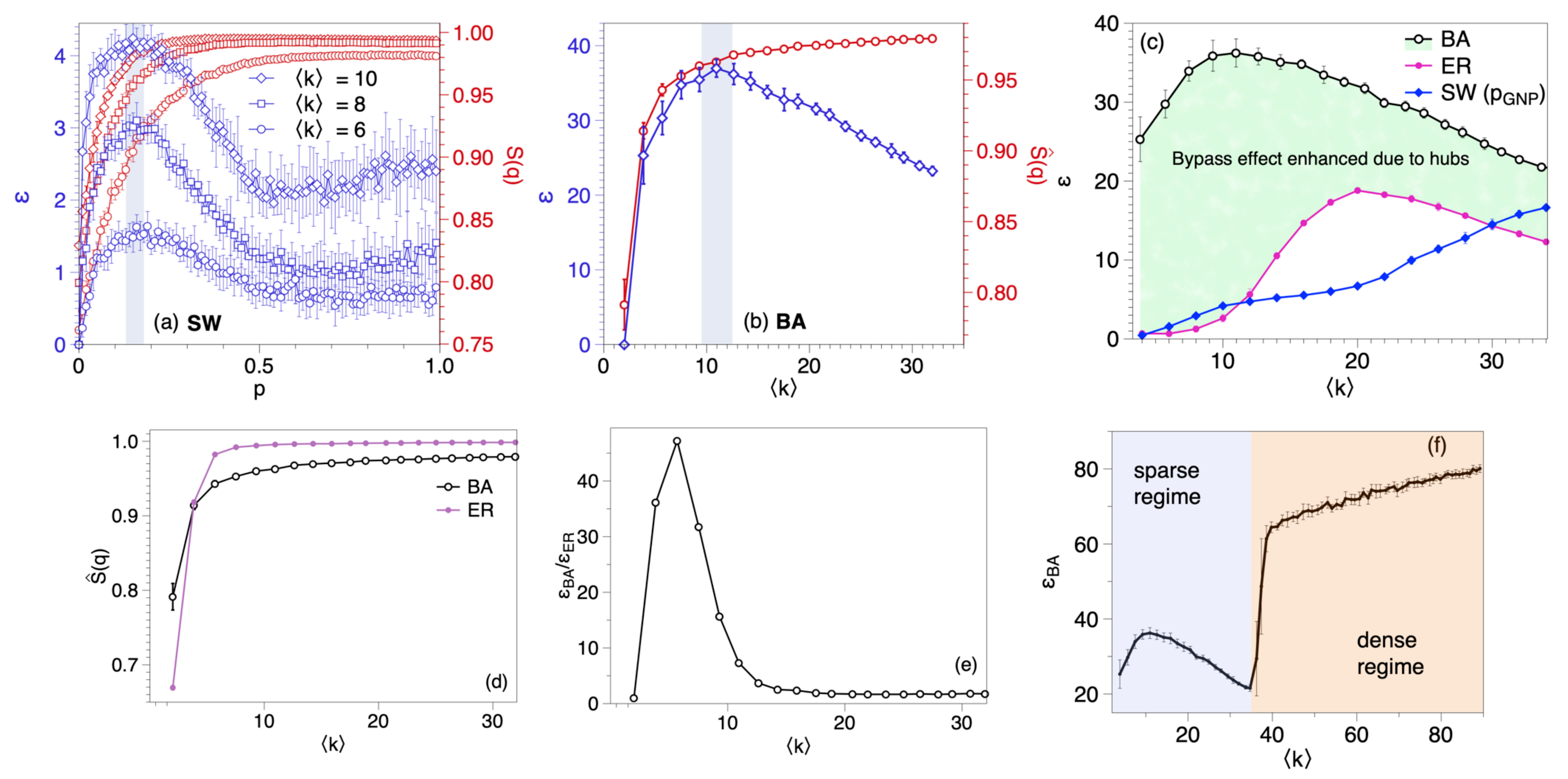}
\par\end{centering}
\caption{Plot of the normalized communicability entropy $\hat{S}({\bf q})$ (red) and of the net gain factor $\epsilon$ (blue)
vs. : ({a}) the rewiring probability $p$ for WS networks (numerical step of $\delta p=0.01$) with $n=250$
nodes and different average degree $\langle k\rangle $, or (b) the mean degree $\langle k\rangle $ of a BA model with $n=250$ nodes. Each dot is the average of 100 realizations, and standard deviations over the ensemble of realizations are also depicted. In both panels, the shaded blue area highlights the maximum of $\epsilon$ and marks the network's Good Navigational Point $(p_{\text{GNP}}\approx 0.15$ for WS model, $\langle k\rangle_{\text{GNP}}\approx 11$ for BA model). {(c)} $\epsilon$ vs $\langle k \rangle$ for networks of $n=250$ nodes generated via the BA model, the WS model (poised at the Good Navigational Point) and an Erdos-Renyi (ER) model for comparison. 
The bypass-induced navigability gain is substantially larger in heterogeneous (BA) networks than in more homogeneous ones (the ratio $\epsilon_{BA}/\epsilon_{ER}$ is plotted in  (e) to highlight such difference). The comparison between ER and SW networks is nontrivial and can be explained in terms of the shapes of the respective degree distributions as $\langle k \rangle$ increases (see the text and SI). {(d)} Normalized communicability entropy of both BA and ER networks with the same number of nodes, as a function of the mean degree.
(f) $\epsilon_{BA}$ vs $\langle k \rangle$ in the extended region of high density, where preferential attachment is not properly working anymore (see SI S9), leading to an explosion of the navigability gain due to the transition to ultra-short graphs. }
\label{WS_bypass}
\end{figure}

\subsection*{Communicability Entropy}

To address the first question above we now quantify, both microscopically and then at the network level, the degree by which, as disorder increases, new routes between edges become available. To this aim, let us return to the WS and BA models that we have considered before. As we have discussed, both the rewiring process and the BA mechanism create degree heterogeneities that intuitively make some a priori `inefficient' paths --e.g. long ones-- to scale up in a pre-defined efficiency ranking (that would indeed be the case of path $P_2$ connecting nodes $i$ and $j$ in Fig.~\ref{WS_sketch_both}).  Now, in practice both WS and BA mechanisms can have heterogeneous effects on this re-ranking, depending on the particularities of the starting and ending nodes $i$ and $j$ (see SI section S3 for an in-depth microscopic analysis on the effect of these local mechanisms on $\xi_{ij}$ and $\mathbb{L}_{\mathfrak{p}(i\to j)}$) . We first start by quantifying how these mechanisms generate a richness of possible trajectories connecting any pair of nodes $i$ and $j$. The probability that a randomly intercepted trajectory indeed corresponds to one connecting $i$ and $j$ is
\begin{equation}
q_{ij}=\frac{G_{ij}}{\sum_{k<l}G_{kl}}.
\end{equation}
Then the heterogeneity in the different number of choices for the trajectory of a particle, {\em i.e.} the trajectory richness of the network is given by the entropy
\begin{equation}
 S({\bf q})=-\frac{1}{2}\sum_{i<j}q_{ij}\ln q_{ij},
 \end{equation}
that we call the communicability entropy. From an information-theoretic perspective, this entropy is a measure of the ignorance we have on who is the sender node and receiver node, when intercepting a message navigating the network. Since $0\leq S({\bf q})\leq \ln(n(n-1)/2)$, the upper bound only reached when the set of probabilities ${\bf q}$ are uniform, we define a normalized version $\hat{S}({\bf q})\coloneqq S({\bf q})/\ln(n(n-1)/2)$.\\ 

\noindent Let us now analyze how $\hat{S}({\bf q})$ behaves in our two reference frameworks. Intuitively, for a fixed mean degree $\langle k \rangle$, $\hat{S}({\bf q})$ will increase in the WS model as $p$ increases, since rewiring increases trajectory richness. Likewise, in a BA model one can vary the network's mean degree: for very small $\langle k \rangle$ the resulting BA network is almost tree-like, with no potential bypasses and thus low trajectory richness, whereas when we allow $\langle k \rangle$ to increase, additional routes are formed thus increasing the trajectory richness, hence $\hat{S}({\bf q})$ should also increase. Figures~\ref{WS_bypass}.a-b (red axis) confirm our intuitive arguments. In particular, in Fig.~\ref{WS_bypass}.a we observe that entropy grows rather quickly in a WS model for small rewiring probability $0<p\leq0.4$, reaching a steady maximum afterwards. The impact of rewiring is notably stronger for small $p$, and this effect is emphasized further for SW networks of increasing $\langle k\rangle$. This behavior is easy to understand: in the small $p$ region there are few shortcuts and each new one makes a difference. On the contrary,  for large values of $p$ the entropy saturates very quickly to $\hat{S}({\bf q})\simeq 1$, i.e. the addition of more shortcuts does not make much of a difference beyond a certain $p$ (see below for further analysis on the influence of the average degree). Figure~\ref{WS_bypass}.b reveals a similar behavior of $\hat{S}({\bf q})$ for the BA model as the mean degree $\langle k \rangle$ increases (within the sparse regime for the BA preferential attachment mechanism to hold, see below), reaching full trajectory richness very quickly after a sudden increase in the region of small $\langle k\rangle$ values. In short, rewiring an ordered structure and increasing the link density of an heterogeneous network quickly (nonlinearly) boosts the trajectory richness, and thus the amount of potential bypasses to any specific shortest path connecting any pair of nodes.\\

We now need to quantify when some of these new routes actually may become consolidated bypasses to shortest paths, like the situation illustrated in Fig.~\ref{WS_sketch_both}, where a particle traveling between two nodes $i$ and $j$ ``might prefer'' to use $P_{2}$, although being longer (in terms of number of edges to be traversed) than the shortest path $P_{1}$. 

\subsection*{Bypass consolidation and associated navigability gain}

To evaluate the impact of potential bypasses on the actual navigability, we use Eq.~(\ref{bigL}) and consider that, for any pair of nodes $i$ and $j$, the SRP between $i$ and $j$ is a consolidated bypass to the shortest path(s) if the effective length of the SRP is smaller than the effective length of the (potentially many) SPs  (i.e. $\mathbb{L}_{SRP}(i,j) <  \mathbb{L}_{SP}(i,j)$ for all SPs connecting $i$ and $j$). Interestingly, this criterion results to be equivalent to check that $\ell_{SRP(i,j)}>\ell_{SP(i,j)}$ (see SI S4 for details). Once bypass detection is done, we need to quantify its impact. A measure that quantifies the impact of bypasses on the network's navigability is the {\it topological length excess} $\epsilon_{(i,j)}$ 
\begin{equation}
\epsilon_{(i,j)}  = \bigg( 1-\frac{\ell_{\text{SP}(i,j)}}{\ell_{\text{SRP}(i,j)}} \bigg)\cdot 100,
\label{eq:excess}
\end{equation}
which indicates that, for a particle traveling between two arbitrary nodes $i$ and $j$, choosing the consolidated bypass SRP over the SP, while beneficial according to the (hidden) network geometry, leads to an {\it apparent} excess of $\epsilon_{(i,j)}  \%$ from the topological distance travelled via the shortest path. It turns out that Eq.\ref{eq:excess} also quantifies the effective distance per link and the resulting gain of using SRP over SP (see SI S4 for a full derivation of these metrics and their interpretation). To extract a global metric for the whole network, we just average $\epsilon_{(i,j)}$ over all pairs of nodes to define the network navigability gain:
\begin{equation}
\epsilon=\frac{2}{N(N-1)}\sum_{i<j}^{N} \epsilon_{(i,j)}.
\end{equation}
An illustration of these metrics in a toy network is given in SI S5. Observe that $\epsilon$ quantifies an improvement of a function (network navigability) as a result of an innovation (consolidation of bypasses) and is therefore a quantitative proxy of structural deepening.\\

\noindent We can now quantify bypass consolidation and its associated navigability gain on relation to both WS and BA mechanisms. When we apply this formalism to the evolving SW network we obtain the results illustrated in panel (a) of Fig.~\ref{WS_bypass} (left axis). We observe that the navigability gain factor $\epsilon$ exhibits a clear non-monotonic shape as a function of the rewiring probability $p$. In fact, our measure detects a {maximum} for $p\approx0.15$ at which, on average, traveling through the SRP is much more favorable than doing so through the SP. We call this probability the ``good navigational point'' (GNP) of the network, $p_{\text{GNP}}$. It is interesting to observe that $p_{\text{GNP}}$ is a precise location inside the so-called small-world regime, which is independent of the network mean degree $\langle k \rangle$ (anecdotally, this value appears close to  the saturating point of spectral spacing in SW networks \cite{brody1, brody2}).\\

\noindent {Now, note that the SW mechanism consolidates bypasses out of a regular-to-random transition, so comparatively speaking the values of $\epsilon$ should be typically higher in more structured networks --e.g. in networks with fat-tailed degree distributions like the BA model-- where the presence of hubs makes the existence of bypasses even more necessary. This hypothesis is confirmed in Fig.\ref{WS_bypass}.b (right axis), in which $\epsilon$ reaches roughly values one order of magnitude larger in the BA model than those found in a comparable WS model. In this case we observe again non-monotonic behavior of $\epsilon$ with $\langle k\rangle$, displaying a {maximum} close to $\langle k \rangle \approx 11$, i.e. the BA model also has a good navigational point when mean degree is $\langle k \rangle_{\text{GNP}}\approx 11$, where bypassing shortest paths that include hubs is maximally relevant.\\

\noindent To further analyse the impact of bypasses, we now compare the values of $\epsilon$ obtained in a BA model ($n=250$ nodes and mean degree $\langle k\rangle$) against (i) those obtained for an Erd\"os-Renyi (ER) graph with the same $n$ and $\langle k\rangle$ --this latter being a model with the same number of edges but with a homogeneous (Poisson) degree distribution and thus virtually lacking any hubs--, and (ii) those of a WS model with the same $n$ and same $\langle k \rangle$, and poised at $p=p_{\text{GNP}}$. Results are shown in Panel (c) of Fig.\ref{WS_bypass} and certify that, in the sparse regime ($\langle k\rangle < 35$), $\epsilon$ is substantially larger in BA than both ER and SW, i.e. the gain supported by bypasses is considerably more important in heterogeneous networks}, as expected \cite{sparse}. When comparing the behavior of $\epsilon$ in ER vs SW networks (both in principle lacking substantial hubs), we observe an interesting effect: for a range of small mean degrees $\langle k \rangle < 11$, SW networks benefit more from bypasses than ER ones. The opposite is true for an intermediate $11< \langle k \rangle < 30$, and the effect is again changed for very large mean degrees $\langle k \rangle > 30$. This nontrivial behavior can be explained by comparing the degree distributions of both ER networks and SW networks at $p_{\text{GNP}}$ and by realizing the (often overlooked) fact that the degree distribution (in particular, the skewness and kurtosis) of a SW network poised at a fixed $p$  undergoes different shapes as $\langle k \rangle$ increases (see SI section S8 for details). Incidentally, this can also explain why $\hat{S}({\bf q})$ initial increase in SW networks is sharper for larger $\langle k \rangle$ (see SI).

\noindent In summary, the effect of bypasses is maximised for SW networks at the Good Navigational Point $p_{\text{GNP}}\approx 0.15$, and within that point, this effect appears to be monotonically boosted when these SW networks increase their degree heterogeneity, i.e. increasing $\langle k \rangle$. ER networks have bypassing properties as long as they show degree heterogeneities, and to a small extent (Poisson distribution) this is the case. Such effect is then maximal around $\langle k \rangle \approx 20$ (the fact that bypasses have a non-monotonic effect also within ER networks can again be explained in terms of the skewness degree distribution, see SI). Finally, in BA networks, bypassing effects are substantially larger due to higher degree heterogeneities, as expected.\\ 

\noindent To close this analysis, one can ask about the theoretical upper bound on $\epsilon$. Heuristically, the effect of bypasses would be fully maximized in a situation where we add to a given (connected) network a new node that is linked to every other node. Such new node would be a `super-hub' that makes the network have shortest paths of length $\leq 2$ for all pairs of nodes. In this extreme situation, many of the shortest paths will be systematically bypassed, and $\epsilon$ would explode (see SI S9 for details). Now, is this just a theoretical scenario? It turns out that this situation can take place in an extreme version of the BA model in a finite graph,  where $\langle k\rangle$ is large enough (compared to the initial seed) so that new nodes entering systematically connect to a large portion of the network, leading to so-called ultra-short graphs \cite{ultrashort}. This explosion is reported in Panel (f) of Fig.\ref{WS_bypass}. Evidently, in this case there is no preferential attachment anymore, so in some sense the rationale behind the BA model breaks down in this dense regime\footnote{In the dense regime $\langle k \rangle >35$  the calculations of $\epsilon$ need to be taken with caution as numerical rounding effects might become important when computing $\exp\left(A\right)$.}.

\begin{figure*}[t!]
\begin{centering}
\includegraphics[width=0.9\textwidth]{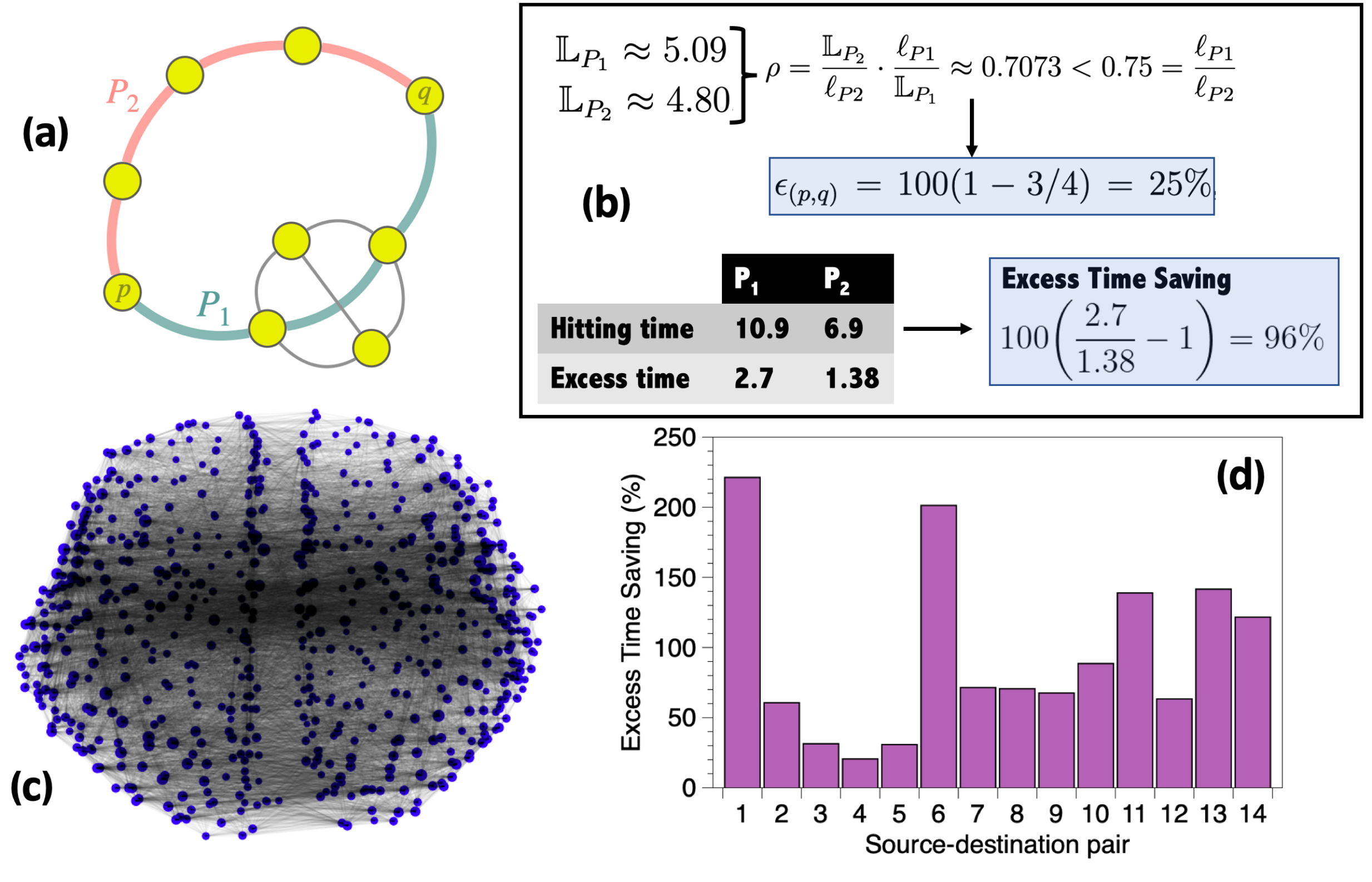}
\par\end{centering}
\caption{{(a) Toy network with a concrete source-destination pair $(p,q)$ which can be navigated via paths $P_1$ and $P_2$. (b) Computation of the different metrics certify that $P_2$ is a bypass of the shortest path $P_1$, and the navigability gain associated to this pair is $25\%$ (a lower bound of the actual gain, see SI S5). Random walk trajectories starting at $p$ and eventually hitting $q$ can be classified as $P_1$-like or $P_2$-like, depending on their specific trajectories (SI S6). The hitting time and excess time of the $P_1$ class is larger than for $P_2$, meaning that random-walk navigability is enhanced in the $P_2$ (that is, the SRP) class. The excess time saving is a diffusion-proxy of the navigability gain (see SI S6). (c) Coactivation brain network. (d) Excess time saving for several source-destination pairs in the brain network, finding that SRP enhances navigability in all cases.}}
\label{fig:Fig3}
\end{figure*}

\subsection*{Effect on Dynamics}

{As already anticipated, our theory is purely structural and therefore dynamically-agnostic, and speaks of the effect of network geometrization on the formation of shortest paths in the geometrized network --the SRPs-- which are different from the shortest paths of the original, ungeometrized network. Our contention is that these emergent bypasses have an effect on the network's navigability, and here we provide a first validation of this hypothesis by considering source-destination random walk trajectories navigating a network. Each of these random walk trajectories is then classified as SRP-like or SP-like depending on the specific sequence of nodes the walker is visiting (see SI S6 for details). One can subsequently compare the SRP class and the SP class by computing a number of quantities, such as the average hitting time in each class, or the excess time (i.e., for each class, how much more time than the time spent by a ballistic walker it takes to reach the destination), which yield dynamical proxies for the effective length or the associated navigability gain defined above (see SI S6). Results for both a synthetic small network and for a large, real network (a coactivation brain network, see below) are shown in Fig.\ref{fig:Fig3}, and confirm our hypothesis that particles are more prone to `get lost' (and thus spend a significantly longer time) navigating through a SP-like path compared to a SRP-like one. In other words, the presence of SRPs enhances navigability for diffusion-like dynamics (additional details and analysis are provided in SI S6). At the same time, this finding further confirms that bypasses induce structural deepening by increasing the efficiency of network navigability.\\

\noindent We have also made some preliminary progress on analysing how bypasses impact other network functions by considering two additional dynamical processes running on a network: synchronization and epidemic spreading. Results (fully detailed in SI section S7) suggest that the prototypical dynamical fingerprints in each case (i.e. eigenratio of the Laplacian matrix for synchronization, and epidemic threshold for epidemic spreading) are affected by bypass consolidation, and in particular, qualitative dynamical changes occur in both types of dynamics close to $p_{\text{GNP}}$.}

\begin{table*}[t!]
\begin{centering}
\begin{tabular}{|c|c|c|c|c|}
\hline 
{\bf Network} & $\hat{S}(q)$ & $ \epsilon  $ (\%) & $p^*$ & $\epsilon/\epsilon_{BA}$\tabularnewline
\hline 
\hline 
Human brain (functional, task-driven) & 0.9234 & 51.71 & 0.30 & 0.78 \tabularnewline
\hline 
Collaboration CoGe & 0.7776 & 41.50 & 0.21 &  0.93 \tabularnewline
\hline 
Collaboration QcGr & 0.4598 & 38.39 & 0.15 &  0.79 \tabularnewline
\hline 
Human brain (anatomical) & $0.925\pm0.022$& $36.21\pm1.52$ & $0.23\pm0.01$ & $0.86\pm0.04$ \tabularnewline
\hline 
\textit{C. elegans} neurons & 0.9312 & 31.69 & 0.34 & 0.87 \tabularnewline
\hline 
USA airports 1997 & 0.8501 & 28.60 & 0.27 & 0.76 \tabularnewline
\hline 
Internet AS 1997 & 0.8891 & 25.49 & $\sim 1$ & 0.52\tabularnewline
\hline 
Yeast PPI & 0.8344 & 25.50 & $0.24$ & 0.55 \tabularnewline
\hline 
Drugs users & 0.7794 & 21.18 & 0.10 &0.57 \tabularnewline
\hline 
Software & $0.8308\pm0.0263$ & $21.11\pm12.10$ & $\sim 1^\dagger$ & $0.58$\tabularnewline
\hline 
Human brain (functional, resting-state) & $0.758\pm0.054$ & $20.81\pm1.42$ & $0.17\pm0.02$ &$0.49\pm0.05$ \tabularnewline
\hline
Roget thesaurus & 0.9215 & 19.18 & 0.35 & 0.43\tabularnewline
\hline 
Transcription yeast & 0.8128 & 12.26 & $\sim 1$ & 0.38\tabularnewline
\hline 
Food webs & $0.9498\pm0.0208$ & $9.94\pm7.13$ & {**} & $0.64^{***}$\tabularnewline
\hline 
electronic circuits & $0.8202\pm0.0260$ & $3.456\pm2.561$ & $\sim 1$ & 0.12\tabularnewline
\hline 
Termite mounds & $0.5707\pm0.0331$ & $3.100\pm2.12$ & $\sim 1$ & 0.11\tabularnewline
\hline 
Power grid & 0.6348 & 2.61 &$\sim 1$ & 0.05 \tabularnewline
\hline 
\end{tabular}
\par\end{centering}
\caption{{Summary of metrics for empirical networks}, depicting the communicability entropy $\hat{S}(q)$, the navigability gain $\epsilon$, the optimal rewiring probability $p^*$ and the navigability ratio $\epsilon/\epsilon_{BA}$ (see the text) across 177 different empirical networks (for many of them, we offer averages, see S10 for details), where: 
{\it Human brain (anatomical)} provides the averaged results across 70 anatomical networks (using the same parcellation), {\it Human brain (functional, resting-state)}
provides the averaged results across 70 functional networks (using the same parcellation as the anatomical networks),
{\it Software} provides the averaged results across the networks MySQL, XMMS, Abi, Digital and VTK; {\it Food
webs} is the average of 15 food webs (see SI S11.3 for disaggregation); {\it Electronic circuits} is the average
of three electronic circuits; {\it Termite mounds} is the average of three
termite mounds. {$\dagger$}Except MySQL which has $p^*\approx0.29$. {*}{*}Three
types of behaviors: (i) $p^*\approx1$ for 8 food webs; (ii) $0.43\protect\leq p^*\protect\leq0.45$
for El Verde, Shelf, Ythan1 and Ythan2; (iii) $0.03\protect\leq p^*\protect\leq0.14$
for Bridge Brooks, Coachella, Little Rock. {***}  See SI section S11 for disaggregated data and additional details.}
\label{table:1}
\end{table*}

\subsection*{Empirical networks}

\begin{figure*}[t!]
\begin{centering}
\includegraphics[width=0.9\textwidth]{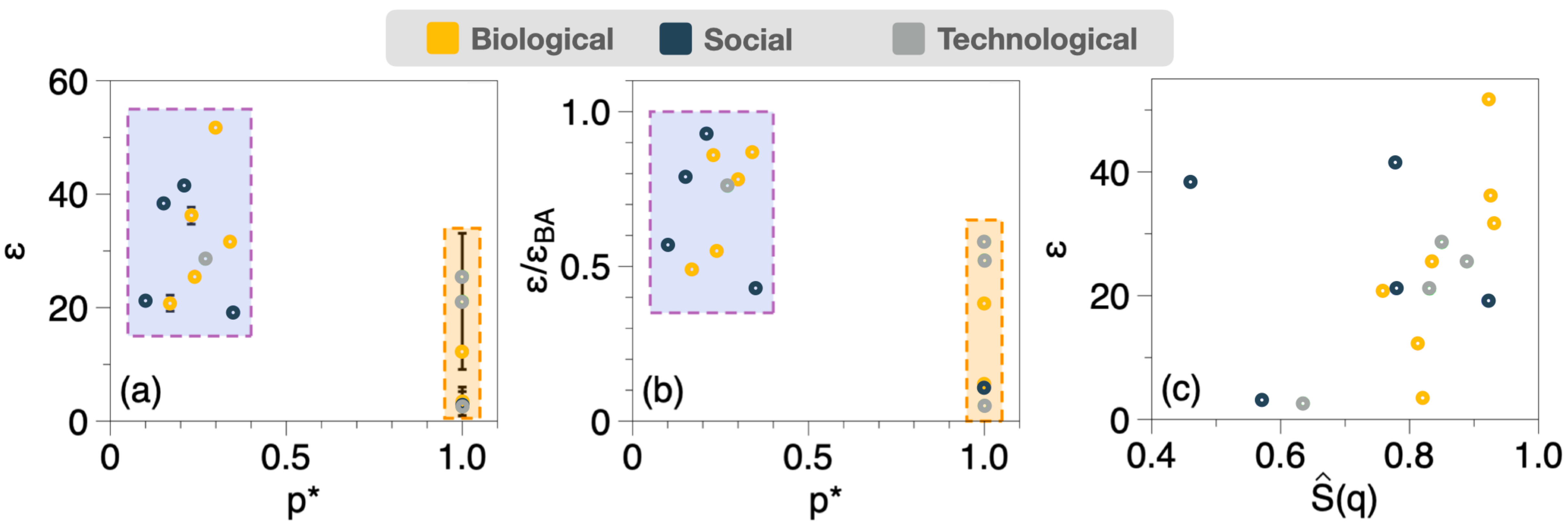}
\par\end{centering}
\caption{{Some scatter plots of the metrics reported in table\ref{table:1} for empirical networks. (a) Net navigability gain $\epsilon$ vs $p^*$, revealing the emergence of two well-defined groups of networks. (b) Navigability ratio $\epsilon/\epsilon_{\text{BA}}$ vs $p^*$, finding the same clustering as in panel (a). (c) $\epsilon$ vs  the normalized communicability entropy $\hat{S}({\bf q})$, where no clear clustering emerges.}}
\label{fig:Fig4}
\end{figure*}

To round off, we have considered a total of {177 empirical networks of different nature, including social (4 collaboration networks of different nature, 3 termite mounds), biological (Human brain --70 anatomical, 70 functional at resting-state, one functional at task-driven (extracted and averaged from a meta-analysis of 1600 works)--}, neural network of \textit{C. elegans}, a protein-protein interaction, a transcription yeast, 15 food webs)  and technological ones (air transportation, Internet, 3 electronic circuits, power grid, 5 software networks), see SI section S11.1 for details and full references. Results on several metrics are summarised in table \ref{table:1} and some scatter-plots are visualised in Fig.\ref{fig:Fig4}.\\
The first two columns of this table report the normalized communicability entropy $\hat{S}(q)$ and navigability gain $\epsilon$. Interestingly, all of them appear to be entropic enough for potential bypasses to have been formed, as values of $\hat{S}(q)$ are in the region where our analysis on synthetic models show consolidated bypasses\footnote{Note, however, that a finer analysis is needed as e.g. we have observed in the SW analysis that reaching the GNP is density-dependent, i.e. the communicability entropy saturates quicker as $p$ increases for networks with larger mean degree.}. We indeed find that essentially all real-world networks harbor consolidated bypasses ($\epsilon >0$), albeit with different impacts, what allows us to rank them accordingly. 
At the top of the ranking, the net gain induced by consolidated bypasses reaches over 50\% for (task-driven) functional brain network, followed by many other self-organised networks (collaboration networks, \textit{C. elegans}, etc). {It is interesting to see that the navigability gain substantially drops for functional brain networks when passing from task-driven activation to resting-state. This might be suggesting the possibility that navigability gain in functional brain networks might be task-related, something that deserves further research. Finding anatomic networks somehow interpolating task-driven and resting-state functional networks is reasonable, given that resting-state function in adults is usually thought to be restricted to a brain module and that the specific task-driven network we analyse is the outcome of a meta-analysis of over 1,600 works considering different tasks --and thus, in principle, multiple brain modules--. These hypothesis await confirmation and, in any case,  further research is needed to elucidate the relation of the topology-induced bypasses studied here with specific cognitive aspects.}\\

\noindent At the bottom of the list in table \ref{table:1} we find some designed networks, such as electronic circuits or the power grid, the latter having only a discrete $2.6\%$ navigability gain. This can be indicative that the power grid, while having hubs to some extent \cite{BA,PG}, has not evolved according to mechanisms such as WS or BA, is not self-organised and as a consequence does not hold the necessary preemptive structural bypasses to avoid systemic failures, as we have seen during blackouts \cite{blackout}. {Note at this point that the navigability gain $\epsilon$ does not trivially correlate with more standard network metrics, such as network density (linear regression of the scatter plot offers a $R^2=0.12$), mean degree ($R^2=0.09$), average path length ($R^2=0.01$) or average clustering ($R^2=0.006$), see SI 11.4 for details.}\\

\noindent Now, to which extent the observed bypasses are indeed of the SW-type (i.e. bypassing shortest paths consistently generated via a WS-like mechanism), and in such case, how close empirical networks are to their theoretical Good Navigational Point? While this question is difficult to answer, the metric $p^*$ reported in the third column of table \ref{table:1} (see also Fig.\ref{fig:Fig4}) provides a first step. Operationally, for a given empirical network $\cal G$ with $n$ nodes and mean degree $\langle k \rangle$, we estimate the closest purely SW-generated network ${\cal G}'(p)$ (with the same $(n, \langle k \rangle)$). This is achieved by minimizing the spectral dissimilarity distance ${\cal D}({\cal G},{\cal G}')=\sqrt{\sum_{j=1}^{n}\left(\lambda_{j}\left({\cal G}\right)-\lambda_{j}\left({\cal {\cal G}}'\right)\right)^{2}}$, where $\lambda_j({\cal G})$ is the $j$-th eigenvalue of the adjacency matrix of network ${\cal G}$ and minimization is over $p$, i.e. $p^*=\text{argmin}_p [{\cal D}({\cal G},{\cal G}'(p)]$\footnote{Note that this analysis is not designed to find which real-world networks can be classified as small-world. It just assumes they all have such ingredient in their network formation, and evaluate, setting that prior as the sole generating mechanism, what would then be the value of the rewiring probability, so as to establish whether the bypass amount is close or not to the GNP. For those networks whose $p^*$ is found to be close to the GNP, this is partial evidence that such network harbors bypasses of the SW-type.}. 
This metric indicates that networks can be typically clustered in two types: one (which includes all human brain networks, the neural network of \textit{C.elegans}, the protein-protein interaction network, collaboration networks, Roget network and the US air network) where ${\cal D}({\cal G},{\cal G}'(p))$ has a non-monotonic shape with a minimum $p^*\in [0.15,0.35]$ --i.e., close but not exactly at the Good Navigational Point--, and another cluster of networks (including electronic circuits, Internet AS97, software networks of termite mounds) where  ${\cal D}({\cal G},{\cal G}'(p))$ is monotonically decreasing and thus $p^*=1$ (see SI S11.2 for further details and analysis). The former class thus tends to harbor bypasses of the SW-type --avoiding shortcuts-- and its network formation includes  at least partially some SW ingredient while the second one tends to have a structure which cannot be well-explained only by SW mechanisms (this doesn't mean, however, that such network is random). Incidentally, no clear function-related clustering emerges.\\

The fourth column of table \ref{table:1} finally depicts $\epsilon/\epsilon_{BA}$ --where $\epsilon_{BA}$ is the navigability gain of a BA network with the same number of nodes $n$ and mean degree $\langle k \rangle$ of the real network--, and quantifies whether the observed network bypasses are effectively bypassing hubs. 
This metric highlights a group of networks where hubs are relevant (i.e. $\epsilon/\epsilon_{BA} \sim 1$) and another group of networks where consolidated bypasses are either not necessarily bypassing hubs, or they are networks that even if having hubs, have not been designed to harbor bypasses and thus do not abide to a BA-like mechanism, so $\epsilon/\epsilon_{BA}$ closer to zero (see SI S11.3 for further discussion).

\section{Discussion}

The journey of network complexification is supported by basic mechanisms including the celebrated WS and BA, among others. As the network evolves accordingly,  we have shown that it naturally increases its communicability entropy $S({\bf q})$ and, in so doing, it allows for new navigational routes to be built, entropically providing bypass `candidates' to the network. Our theory allows to detect when some of these new routes consolidate their bypassing property by subsequently getting to be more favorable than the corresponding shortest paths connecting the same pairs of nodes, and we show that consolidation takes place in both WS and BA models. Interestingly, we find that the role of bypasses is maximised in a small parameter region --which we call the network's Good Navigational Point--, located in a point inside the Small-World regime and for a specific mean degree in the BA model. These findings suggest that the navigation gain offered by the network bypasses is indeed reflecting a form of structural deepening, thus putting the onset of complexity in networks into a solid quantitative footing.\\

\noindent We have certified that bypasses induce clear navigation gain for particles undergoing diffusion-like dynamics and also play an effect on other network functions, including harboring synchronization and epidemic spreading. We have then shown that many empirical networks considered complex, including brain networks, indeed have Good Navigational Point properties, while those that are not catalogued as self-organised but have been designed tend to not include bypasses in their design, with well-known unfortunate consequences \cite{blackout}.\\

\noindent In hindsight, our results could provide a theoretical and mechanistic support for the role of bypasses in e.g. physiological systems, --where plasticity is of utmost importance \cite{gosak}--. First, network bypasses naturally relate to the existence of the so-called \textquotedblleft collateral
circulation\textquotedblright: a system of specialized endogenous bypass vessels present in most tissues providing protection against ischemic injury caused by ischemic stroke, coronary atherosclerosis, peripheral artery disease and other conditions and diseases \cite{circulation}.
Second, in brain networks,
there is nowadays enough observational evidence which supports that these are SW in the Watts-Strogatz sense \cite{BrainSW} and possess hubs which create skewness of their degree distributions \cite{Brainhubs}. At the same time, recent experiments \cite{Energy_cost} suggest that propagating signals in the brain using hubs as part of the navigation path might have a large energetic cost, triggering research on non-geodesic information propagation  \cite{Energy_cost,Navigation_1,Navigation_2}. Our work indeed supports the concept of non-geodesic navigability (via network bypasses), and reconciles this with the reported network structure. In this context, note that \cite{yin} proposed considering networks of neurons as evolving and growing connections in a distributed fashion (via mechanisms different that SW or BA), such that shortest path minimization and robustness maximization (which in general implied to avoid the creation of hubs) was performed at the same time. Note, however, that brain navigation is not likely to occur always geodesically \cite{Energy_cost,Navigation_1,Navigation_2} (
this also would imply that individual neurons perform global optimization and have access to the the whole brain structure). The logical conclusion is that the seminal findings in \cite{yin} imply that the creation of shortest paths should be accompanied by the proliferation of additional structure that plays a role of structural deepening, in good agreement with our theory.\\

\noindent Third, in another recent work \cite{Arenas} it has been shown  that brain function appears to be robust against damage by re-adapting and re-purposing non-damaged links, something that can be interpreted to the brain's ability to re-compute SRPs and thus re-rank bypasses after network damage. All in all, elucidating the impact of our findings in the context of neuroscience is an exciting avenue for future work.\\ 

\noindent An aspect not explored in this paper but also of major interest is the implications of our theory to congestion or jamming phenomena in networks, and to which extent our proposed measures of topological length excess and navigability gain could anticipate congestion in e.g. transportation and urban systems. First, we should disclose that conceptually similar problems have been theorized in the mathematics literature, where some authors have studied the so-called ``resistance distance'' in networks \cite{klein} --where some unit resistances are placed at every edge in a network-- in the context of congestion \cite{grippo,jonckheere}. Now, while an interesting mathematical concept, this latter distance analytically converges, for large graphs and in high dimensions, to an expression that does not take into account the structure of the graph  \cite{Luxburg} (i.e. it only depends on the degrees of the source and destination nodes) and thus unfortunately turns useless in real-world scenarios\footnote{Anecdotally,  it is easy to see that such resistance distance is already unable to capture the nuanced navigability properties of paths P1 and P2 in the toy network presented in Fig.\ref{fig:Fig3} (revealed by the hitting times analysis) and that in turn our theory correctly predicts.}. Second, 
recent empirical evidence in urban science indeed suggests that, at rush hours, in different cities worldwide paths which can be identified as SRPs are supporting more traffic than SPs \cite{Akbarzadeh}, i.e. they become systematically preferred routes. This constitutes preliminary support in favor of the relevance of SRPs for navigation strategies in networks subject to jamming, and further research is deserved. For instance, we speculate that this strategy can be further refined by, instead of systematically selecting only the SRP as the preferred route, ranking each of the paths connecting any two locations via the computation of its associated topological length excess and rerouting traffic accordingly when needed.\\
Other important open questions for further research include understanding the role played by network bypasses and their relation to structural deepening in other mechanistic growth models (e.g. assortative/dissasortative mixing, triadic closure, etc), and the extension of our theory to weighted (see a preliminary discussion on this topic in SI S10), temporal and higher-order networks \cite{hoi, hoi2}.\\

\noindent Finally, while network bypass emergence appears to be contingent on the growth mechanism --and thus appears to be a byproduct of it--,  bypass consolidation (structural deepening) is the effect which probably makes those growth mechanisms to be sustainable in the first place. Simply put, we argue, bypasses sustain complexity.

\paragraph{Acknowledgments}
The authors thank Olaf Sporns, Yasser Aleman-Gomez and Yasser Iturria-Medina for assistance with the data and analysis of brain networks, Adrian Garcia-Candel for help formatting Fig1, and referees for insightful comments. EE acknowledges funding from project OLGRA (PID2019-107603GB-I00) funded by Spanish Ministry of Science and Innovation. 
J.G.-G. acknowledges financial support from the Departamento de Industria e Innovaci\'on del Gobierno de Arag\'on y Fondo Social Europeo (FENOL group grant E36-23R) and from project TEAMS (PID2020-113582GB-I00) funded by the Spanish Ministry of Science and Innovation. L.L. acknowledges funding from project DYNDEEP (EUR2021-122007) and project MISLAND (PID2020-114324GB-C22), both projects funded by Spanish Ministry of Science and Innovation. This work has also been partially supported by the Maria de Maeztu project CEX2021-001164-M funded by the MCIN/AEI/10.13039/501100011033.
\vspace{-5mm}

\paragraph{Contributions}
EE and LL designed the research and EE performed the computational and analytical calculations. LL and JGG contributed additional computational analysis. EE and LL wrote the first draft of the paper. All authors discussed results and revised the paper.\vspace{-5mm}
\paragraph{Competing financial interests}
The authors declare no competing financial interests.
\vspace{-5mm}
\paragraph{Additional information}
Correspondence and requests for materials should be addressed to Ernesto Estrada (estrada@ifisc.uib-csic.es).


\newpage

\section*{Supplementary information}

\subsection*{S1. Network Communicability $G_{ij}(\beta)$}

\paragraph{S1.1 Network communicability across the disciplines.} The communicability function emerges in a variety of physical scenarios:  it was proved in \cite{oscillators} that $G_{ij}\left(\beta\right)$ corresponds to the thermal Green's function of a network of coupled quantum harmonic oscillators where $\beta$ represents the inverse temperature of the thermal bath in which the network is submerged.  The self-communicability $G_{ii}\left(\beta\right)$ also appears in the definition of the probability of finding a network, represented by a tight-binding Hamiltonian, in a state with energy $E_{j}=-\lambda_{j}$ when the system has inverse temperature equal to $\beta$ \cite{subgraph centrality}. More recently, Lee et al. \cite{LTE} found that the communicability function appears in the solution of a linearised upper bound to the susceptible-infected (SI) model when certain initial conditions are impossed to the epidemic dynamics. Similar results were extended by Bartesaghi and Estrada \cite{Bartesaghi} for the SIS model. Finally, the communicability emerges as the solution of a linear autonomous system, which is a modification of the Kuramoto model, when the internal frequency of the oscillators is zero \cite{Arola}. Nowadays communicability is used for the analysis of the structure and dynamics on brain networks, granular material, infrastructural systems, social networks, biological patterns, genomic and proteomic systems, among others (see references in the main article).

\paragraph{S1.2 Derivation of the communicability function in a concrete setting.} For illustration, here we present the derivation of the communicability function and subsequent operators in case for the dynamics of a particle hopping on a network following the tight-binding Hamiltonian:
\begin{equation}
\mathcal{H}\coloneqq\sum_{v=1}^{n}\varepsilon_{v}\left|v\right\rangle \left\langle v\right|+\sum_{v\neq w}^{n}A_{vw}t_{mn}\left|w\right\rangle \left\langle v\right|,
\label{hamil}
\end{equation}
where ${\bf A}=\{A_{vw}\}_{v,w=1}^n$ is the network's adjacency matrix, and $t_{mn}$ is the hopping parameter, which represents the energy that facilitates the hopping of a particle from v to w. If this parameters tends to zero the particles cannot move from one site to another (an isolated or trivial graph). If we turn on the parameter, the particles can hope from one site to the other. Here we fix it as $t_{mn}=1$ for the sake of simplicity. The eigenstates of the Hamiltonian are represented in the Diract bra-kets notation as usual.\\

\noindent In the initial value representation, the probability amplitude for
a transition from state (node) $v$ at time zero to state (node) $w$ at time $t$
is given by $G_{vw}\left(t\right)=\left\langle \psi_{w}\right|e^{-it\mathcal{H}/\hbar}$$\left|\psi_{v}\right\rangle $.
Applying a Wick rotation $t\rightarrow i\beta\hbar$, the real-time propagator $e^{-it\mathcal{H}/\hbar}$ is mapped  into the thermal
propagator $e^{-\beta\mathcal{H}}$, where $\beta$
is the inverse temperature of a thermal bath in which the network
is submerged into. Setting $\varepsilon_{v}=0$ for every
node and $t_{mn}=-1$ for every pair of nodes, as it is done in the tight-binding model, we identify $\mathcal{H}=-{\bf A}$
and the propagator reduces to the communicability function $G(\beta)=e^{\beta {\bf A}}$.\\ 

\noindent Then, for a particle located at the node $v$ at time zero, the difference:
\begin{equation}
R_{vw}\left(\beta\right)\coloneqq\left\langle \psi_{v}\right|e^{\beta {\bf A}}\left|\psi_{v}\right\rangle -\left\langle \psi_{w}\right|e^{\beta {\bf A}}\left|\psi_{v}\right\rangle ,
\end{equation}
accounts for the opposition offered by the network to the displacement of the particle to the node $w$ at inverse temperature $\beta$. The smallest the values of $R_{vw}\left(\beta\right)$ the highest the probability that the particle does not get trapped at the origin and can propagate to node $w.$ One can symmetrize this quantity to define the following operator:
\begin{equation}
\begin{split}\xi^2_{vw}\left(\beta\right) & \coloneqq R_{vw}\left(\beta\right)+R_{wv}\left(\beta\right)=\left\langle \psi_{v}-\psi_{w}\right|e^{\beta {\bf A}}\left|\psi_{v}-\psi_{w}\right\rangle.
\end{split}
\end{equation}
Mathematically, one can prove that $\xi_{vw}$ is an Euclidean distance between the corresponding nodes
(see SI below). {Conceptually, $\xi_{vw}$ is indeed a measure of the network opposition to the flow between $v$ and $w$}. As a matter of fact, and as we will show below, $\xi_{vw}$ is indeed formally equivalent to the so-called effective resistance found between a pair of nodes $v$ and $w$ in an electric circuit, which is expressed as $\langle \psi_v - \psi_w | L^* |  \psi_v - \psi_w \rangle$, where $L^*$ is the Moore-Penrose pseudoinverse of the circuit's Laplacian matrix. By analogy, we call $\xi_{vw}$ the communicability resistance between nodes $v$ and $w$ in the network.

\subsection*{S2 $\xi_{vw}$ is an Euclidean distance}
\begin{theorem}
Let $\xi^2_{vw}\coloneqq\left(e^{A}\right)_{vv}+\left(e^{A}\right)_{ww}-2\left(e^{A}\right)_{vw}$, where $A$ is the adjacency matrix of a graph.
Then, $\xi_{vw}$ is an Euclidean distance between the nodes
$v$ and $w$.
\end{theorem}
\begin{svmultproof}
Let $A=U\varLambda U^{T}$ where the $j$th column of $U$ is $\psi_{j}$
and $\Lambda=diag\left(\lambda_{j}\right)$. Then,
\begin{equation}
\begin{split}\xi^2_{vw} & =\sum_{j=1}^{n}e^{\lambda_{j}}\psi_{jv}^{2}+\sum_{j=1}^{n}e^{\lambda_{j}}\psi_{jw}^{2}-2\sum_{j=1}^{n}e^{\lambda_{j}}\psi_{jv}\psi_{jw}\\
 & =\sum_{j=1}^{n}e^{\lambda_{j}}\left(\psi_{jv}-\psi_{jw}\right)^{2}
\end{split}
\end{equation}
Let us define $\varphi_{v}$ to be the $v$th column of $U^{T}$,
such that we can write 
\begin{equation}
\begin{split}\xi^2_{vw} & =\left(\varphi_{v}-\varphi_{w}\right)^{T}e^{\varLambda}\left(\varphi_{v}-\varphi_{w}\right),\end{split}
\end{equation}
then, because $e^{\varLambda}$ is positive definite we can write
\begin{equation}
\begin{split}\xi^2_{vw} & =\left(e^{\varLambda/2}\varphi_{v}-e^{\varLambda/2}\varphi_{w}\right)^{T}\left(e^{\varLambda/2}\varphi_{v}-e^{\varLambda/2}\varphi_{w}\right).\end{split}
\end{equation}
Let us designate $x_{v}\coloneqq e^{\varLambda/2}\varphi_{v}$, such
that we have
\begin{equation}
\begin{split}\xi^2_{vw} & =\left(x_{v}-x_{w}\right)^{T}\left(x_{v}-x_{w}\right)\\
 & =\left\Vert x_{v}-x_{w}\right\Vert ^{2},
\end{split}
\end{equation}
which means that ${\xi_{vw}}=\sqrt{\left\Vert x_{v}-x_{w}\right\Vert ^{2}}$
is a Euclidean distance between $v$ and $w$. Therefore, $x_{v}$
and $x_{w}$ are the coordinates of the nodes $v$ and $w$ in such
Euclidean space.
\end{svmultproof}

\subsection*{S3 How communicability and effective distance change with network rewiring: a microscopic analysis}

Let us consider the effective length of a path $P$ between the nodes
$i$ and $j$ as
\begin{equation}
\begin{split}\mathbb{L}_{P\left(i,j\right)} & =\sum_{e_{k}\in P}\xi_{e_{k}}=\left(\sum_{e_{k}\in P}\xi_{e_{k}}^{2}+2\sum_{e_{k}\neq e_{l}}\xi_{e_{k}}\xi_{e_{l}}\right)^{1/2},
\end{split}
\label{sqrt}
\end{equation}
where $e_{k}$ is an edge in the path $P$, where the second equality is used for convenience and simply uses the fact that $a+b=\sqrt{a^2 + b^2 + 2ab}$ (which can always be made since $\xi$ is a distance and thus non-negative).\\
The first term inside the square root is given by
\begin{equation}
\begin{split}\sum_{e_{k}\in P}\xi_{e_{k}}^{2} & =G_{ii}+2\sum_{s\in P}G_{ss}+G_{jj}-2\sum_{\left(r,t\right)\in e_{k}\in P}G_{rt},\end{split}
\end{equation}
where $i$ and $j$ are the endpoints of the path $P$, $s$ are intermediate
nodes in the path and $\left(r,t\right)$ are every adjacent pairs
of nodes belonging to the corresponding path.\\ 
Then, in order to drop the effective length of the path $P$ relative to another
path connecting the same pair of nodes $i$ and $j$ --for instance,
relative to the SP-- we can act in any of the following two cases:
\begin{enumerate}
\item Dropping $\sum_{s\in P}G_{ss}$ to any node in $P$ relative to the
similar sum in the reference path. This can be done by dropping the
number of small subgraphs in which nodes in the path $P$ participates.
For instance, we can delete edges or triangles attached to any node
in the path $P$.
\item Increasing $\sum_{\left(r,t\right)\in e_{k}\in P}G_{\left(rt\right)}$
to any pair of nodes forming an edge in $P$ relative to the similar
sum in the reference path. This can be done by increasing the communication
capacity of pairs of nodes in the path $P$. For instance, by dropping
the length of the cycles in which the two nodes are involved or by
increasing the number of paths between pairs of nodes in the path
$P$.
\end{enumerate}
Using only this contribution we can see how the two growing mechanisms
studied in this paper may affect these two cases. The rewiring process
of the WS mechanism can influence the two cases analyzed in the following
way. The rewiring may remove edges attached to nodes which are in
the path under consideration (Case 1), at the same time such rewires
can create shortcuts that increases the communicability between alternative
paths (Case 2). Additionally, the rewiring may attach edges to pairs
of nodes in the SP and also can drop the communicability between others
which are in the SP. In the case of the BA model, in must be remarked
that the addition of a node connected to the nodes of the network
will not drop the self-communicaility of any node (Case 1 is not possible).
Therefore, such mechanism can drop the effective length of a path
by increasing the communicability between pairs of nodes in the path
or by increasing the self-communicability of nodes in the SP.

\noindent The second term in Eq.\ref{sqrt} represents the interaction between pair of edges in
the graph, where each individual term can be written as:
\begin{equation}
\xi_{e_{k}}\xi_{e_{l}}=\left(\sum_{s\in e_{k},r\in e_{l}}G_{ss}G_{rr}-2\sum_{s\in e_{k},\left(r,t\right)\in e_{l}}G_{ss}G_{rt}+2\sum_{\left(p,q\right)\in e_{k},\left(r,t\right)\in e_{l}}G_{pq}G_{rt}\right)^{1/2},
\end{equation}
where the first term is the sum of the product of self-communicabilities
of nodes $s$ and $r$, where the two nodes are always in different
edges, the second term is the product of the self-communicability
of a node in one edge with the communicability between a pair of nodes
in another edge, and the last term is the product of the communicability
between two nodes in one edge and the one between the two nodes in
another.
This second term in the expression of $\mathbb{L}_{P\left(i,j\right)}$ indeed 
introduces nonlinearities in the influence of the self-communicability
and internode communicability on the change of the length of a path
after a given transformation. For, instance, we have seen that dropping
$\sum_{s\in P}G_{ss}$ will make the path less resistant, according
to the first term of the expression. This dropping also contributes
to drop $\sum_{s\in e_{k},r\in e_{l}}G_{ss}G_{rr},$ which makes the
path less resistant, but it contributes to increase the effective
length of the path $P$ via the term $2\sum_{s\in e_{k},\left(r,t\right)\in e_{l}}G_{ss}G_{rt}.$
A similar analysis is possible with the communicability $G_{rt},$
whose increment drops the term $2\sum_{s\in e_{k},\left(r,t\right)\in e_{l}}G_{ss}G_{rt},$
but increases $2\sum_{\left(p,q\right)\in e_{k},\left(r,t\right)\in e_{l}}G_{pq}G_{rt}.$
As a consequence of the different types of contributions emerging
in both terms, there will be a trade off between the self-communibility
and the internode communicability, such that changing them in the
appropriate direction will drop the effective length of the path $P$
relative to the SP only up to certain point after which the effective
length will increase.

\subsection*{S4 Derivation of the bypass consolidation criterion and measures of impact of bypass consolidation in network navigability}

\paragraph{S4.1 Bypass consolidation criterion} Take two arbitrary nodes $s$ and $t$, and denote ${\cal S}_{(s,t)}$ the set of all shortest paths connecting $s$ and $t$ (this set has at least one element, but it might have many more, and that is typically the case when the network is large and the nodes $s$,$t$ are distant, topologically speaking).
Let us denote ${\cal P}_{(s,t)}$ the set of potential bypasses. This consists of all paths connecting $s$ and $t$ except the set of shortest paths $${\cal P}_{(s,t)}=\{\mathfrak{p}(s\to t)\} \setminus {\cal S}_{(s,t)}.$$ We say that any given path $\mathfrak{p}(s\to t)$ is a consolidated bypass if 
\begin{equation}
\mathbb{L}_{\mathfrak{p}(s\to t)} < \min_{\text{SP}(s,t) \in {\cal S}_{(s,t)}}[\mathbb{L}_{\text{SP}(s,t)}].
\label{eq:criterion_all}
\end{equation}
It is clear that, for a given pair of nodes $(s,t)$, applying Eq.\ref{eq:criterion_all} for all elements in ${\cal P}_{(s,t)}$ can yield no solutions, one solution, or multiple solutions. To make the quantification simpler, instead of looking for the total number of consolidated bypasses of $(s,t)$, we only consider a specific path, i.e. the one which has the smallest effective length, i.e. the SRP. In this way, we say that the shortest path(s) are bypassed if
\begin{equation}
\mathbb{L}_{SRP(s,t)} < \min_{\text{SP}(s,t) \in {\cal S}_{(s,t)}}[\mathbb{L}_{\text{SP}(s,t)}].
\label{eq:criterion_SRP}
\end{equation}
Observe that, by definition of the Shortest Resistive Path, for any shortest path we always have
$$\mathbb{L}_{SRP(s,t)} \leq \mathbb{L}_{SP(s,t)},$$
and at the same time, by definition of the Shortest Path(s), we always have
\begin{equation}
\ell_{SP(s,t)} \leq \ell_{SRP(s,t)}.
\label{eq:SP_condition}
\end{equation}
These two properties imply that if Eq.\ref{eq:criterion_SRP} is satisfied, then necessarily we have as a consequence that 
\begin{equation}
\ell_{SRP(s,t)}>\ell_{SP(s,t)}.
\label{eq:crit_feten}
\end{equation}
In fact, if 
$\ell_{SRP(s,t)}>\ell_{SP(s,t)}$ then it follows that $\mathbb{L}_{SRP(s,t)} < \min_{\text{SP}(s,t) \in {\cal S}_{(s,t)}}[\mathbb{L}_{\text{SP}(s,t)}]$ (the other alternative would imply that the SRP is indeed a shortest path and thus not a bypass, resulting in a contradiction). Accordingly, SRP$(s,t)$ is a consolidated bypass if and only if Eq.\ref{eq:crit_feten} holds. In fact the same criterion holds to assess if any ${\mathfrak{p}(s\to t)}$ is a consolidated bypass, but as previously mentioned we won't be analysing the abundance of bypasses, only whether for a given $s,t$, one consolidated bypass exists. It is trivial to see that checking the SRT is a sufficient condition, so we focus on the SRT.

\paragraph{S4.2 Quantification of bypass effect} Once we have a principled way of assessing if the SRP is a consolidated bypass, we wish to characterize its effect on the navigability of the network. There are two independent aspects one can look at.\\
First, one can assess how larger is the topological distance of the SRT with respect to the SP, i.e. how small the ratio
$\frac{\ell_{SP(s,t)}}{\ell_{SRP(s,t)}}<1$ is. From this we can define the {\it topological length excess} $\epsilon_{(s,t)}$ 
\begin{equation}
\epsilon_{(s,t)}  = \bigg( 1-\frac{\ell_{\text{SP}(s,t)}}{\ell_{\text{SRP}(t,t)}} \bigg)\cdot 100,
\label{eq:excess}
\end{equation}
which indicates that, for a particle traveling between two arbitrary nodes $i$ and $j$, the choice of consolidated bypass SRP over the SP, while beneficial, leads to an {\it apparent} excess of $\epsilon_{(s,t)}  \%$ from the topological distance travelled via the shortest path.
Since Eq.\ref{eq:excess} is a ratio, then it can be properly averaged across all pairs to construct an average topological length excess
\begin{equation}
\epsilon=\frac{2}{N(N-1)}\sum_{s<t}^{N} \epsilon_{(s,t)} ,
\label{eq:excess_all}
\end{equation}
that quantifies the average effect of bypasses across all pairs of nodes in terms of the resulting increase in topological length.\\
Second,  we can ask ourselves why would a walker prefer to take a path of, say, topological length $\ell=9$, than another of, say, topological length $\ell=3$? Of course this is the case if the first path has smaller effective distance (the walker might diffuse out in the second route, thus getting lost --if the walker was a diffusive one-- using that other route, etc), but this information does not seem to be easily interpreted from Eqs.\ref{eq:excess} and \ref{eq:excess_all}. So to some extent knowing that a walker still prefers a path of topological length three times larger gives also rough idea of the underlying (hidden) resistive geometry. Let us explain this:
For the sake or argument, let us suppose that we have detected a consolidated bypass SRP with a certain effective length $\mathbb{L}_{SRP}<\mathbb{L}_{SP}$, and consider two hypothetical situations: (1) in the first one we would have $\ell_{SRP(s,t)} \gg \ell_{SP(s,t)}$, and (2) in the second, $\ell_{SRP(s,t)}$ is only marginally larger than $\ell_{SP(s,t)}$.  Clearly, in the first situation the walker navigating from $s$ to $t$ choosing SRP would need to traverse a proportionally much larger amount of edges than by choosing the SP than in the second situation. Since, we need to recall, this is utterly beneficial (because SRP is at the end of the day a consolidated bypass of the SP), this necessarily means that not only the effective length of the SRP is {\it smaller} than the effective length of the SP (true by definition of the consolidated bypass), but furthermore, it means that the ratio between the effective length {\it per edge} in the SRP and in the SP is notably smaller in the first situation than in the second. In formulas, the average effective length per edge in the SRP and SP are
$$ \langle\mathbb{L}_{SRP(s,t)} \rangle = \frac{\mathbb{L}_{SRP(s,t)}}{\ell_{SRP(s,t)}}; \ \ \langle\mathbb{L}_{SP(s,t)} \rangle = \frac{\mathbb{L}_{SP(s,t)}}{\ell_{SP(s,t)}}$$
and then the ratio
\begin{equation}
\rho=\frac{\langle\mathbb{L}_{SRP(s,t)} \rangle}{\langle\mathbb{L}_{SP(s,t)} \rangle}
\label{eq:rho}
\end{equation}
quantifies the net benefit (reduction) in effective distance per edge in choosing SRP over SP. $\rho$ is thus the correct measure to compare (and also to average) such benefit for different bypasses. 
{Unfortunately, evaluating Eq.\ref{eq:rho} is only computationally efficient for small networks, and in general is not scalable (see however next section for a concrete illustration where such computation is performed). This is mainly due to the fact that in a large unweighted network, in general between any two nodes there might be very many shortest paths --all of them with the same topological length but different effective length--.  To exhaustively enumerate all these paths is in general a difficult task, and it is thus much easier to compute $\ell_{SP(s,t)}$ than $\mathbb{L}_{SP(s,t)}$. On the other hand, computing $\mathbb{L}_{SRP(s,t)}$ is actually computationally efficient: once the network is geometrized, it effectively converts into a weighted network with real-valued weights, and the shortest path in this geometrized network (which is indeed the SRP of the original network) is almost surely unique, so a Dijkstra's algorithm finds it without ambiguity, and without the needs to exhaustively enumerate all possible paths, including the very many shortest paths of the original network.} 
Interestingly, by using the properties of SRP and SP we can show that 
\begin{equation}
\rho = \frac{\mathbb{L}_{SRP(s,t)}}{\mathbb{L}_{SP(s,t)}}  \frac{ \ell_{SP(s,t)}}{\ell_{SRP(s,t)}}\leq \frac{\ell_{SP(s,t)}}{\ell_{SRP(s,t)}}\leq 1.
\label{bound}
\end{equation}
In words, $\rho$ quantifies how much more beneficial it is --in terms of effective distance travelled per edge-- to travels through the SRP than through the SP, and this gain is bounded by  $\ell_{\text{SP}(s,t)}/\ell_{\text{RSP}(s,t)}$. This upper bound is computationally efficient to work out for large networks and in fact, we have already computed it in the bypass consolidation criterion and in Eqs.\ref{eq:excess} and \ref{eq:excess_all} ! Using this upper bound as a conservative approximation of $\rho$, it turns out that the {\it topological length excess} $\epsilon_{(s,t)}$ can now be interpreted as a lower bound of the {\it effective distance gain per edge} of using SRP over SP, and likewise $\epsilon$ is the network's {\it average effective distance saving} in the navigability that bypasses have globally provided.

\begin{figure}[htb]
\begin{centering}
\includegraphics[width=0.48\textwidth]{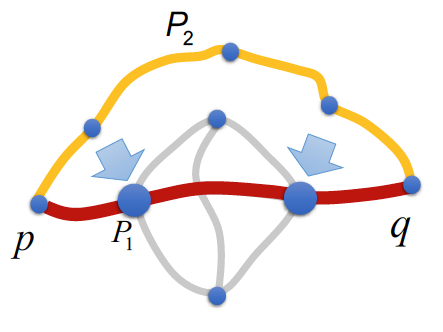}
\par\end{centering}
\caption{{\small{\bf Explicit example} Cartoon network where $P_2$ (highlighted in yellow) is a consolidated bypass of the shortest path $P_1$ (highlighted in red) since $P_1$ has smaller effective length than $P_2$, despite of having a longer topological length. This happens because the shortest path involves some nodes with higher degree, which are sources where a walker could diffuse out.}
\label{fig:cartoon}}
\end{figure}

\subsection*{{S5 Validation on a toy network}}

\paragraph{{S5.1 The toy network}} Let us put the formalism in action in a concrete example, a network of $n=9$ nodes illustrated in Fig.\ref{fig:cartoon}. In particular, $P_1$ is the shortest path connecting the nodes $p$ and $q$ (highlighted in red), and has topological length $\ell_{P_1}=3$. This path however includes two `hubs', proned to diffuse out particles (or to be damaged or jammed), and there exists an alternative route ($P_2$, highlighted in yellow). $P_2$ is a potential bypass with larger topological distance $\ell_{P_2}=4$. 

\paragraph{{S5.2 Topological length excess and $\rho$}}
$P_2$ is indeed a consolidated bypass since $\mathbb{L}_{P_1}\approx5.09$ and $\mathbb{L}_{P_2}\approx4.80$, so $\mathbb{L}_{P_2}< \mathbb{L}_{P_1}$: while $P_1$ is topologically shorter, in units of the hidden resistive geometry, it is effectively longer! The topological length excess is
$\epsilon_{(p,q)}=100(1-3/4)=25 \%$, i.e. the walker will appear to be choosing a path with $25\%$ more links than the shortest path. Now, the effective distance per edge ratio 
$$\rho = \frac{\mathbb{L}_{P_2}}{\ell_{P2}} \cdot \frac{\ell_{P1}}{\mathbb{L}_{P_1}} \approx 0.7073 < 0.75 = \frac{\ell_{P1}}{\ell_{P2}}.$$ 
In words, by actually choosing $P_2$, the walker will only travel in each link a distance which is about $70\%$ of the actual (effective) distance travelled via $P_1$. This is less than $75\%$, the bound offered by the topological ratio. So the net saving of taking $P_2$ instead of the shortest path $P_1$ is indeed $100(1-0.7073)\approx 30\%$, but our lower bound offers a conservative $\epsilon = 25\%$.\\
Putting this example in a general context, this helps to illustrate the fact that by choosing the SRP over the SP, the walkers are reducing the effective length of their walk, therefore enhancing navigability. Our estimation of how much this is enhanced ($\epsilon$) is only a conservative lower bound of the actual number.

\subsection*{{S6 Hitting times and Excess times of SRP-like and SP-like random walkers}}
\begin{figure}[t!]
\begin{centering}
\includegraphics[width=0.48\textwidth]{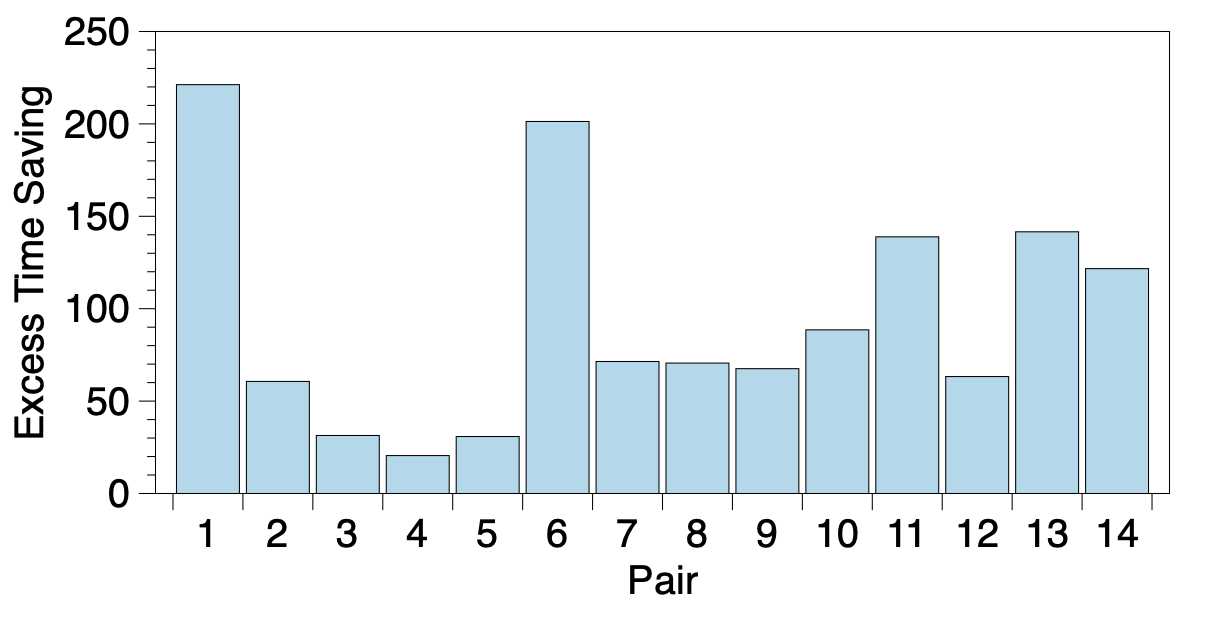}
\includegraphics[width=0.48\textwidth]{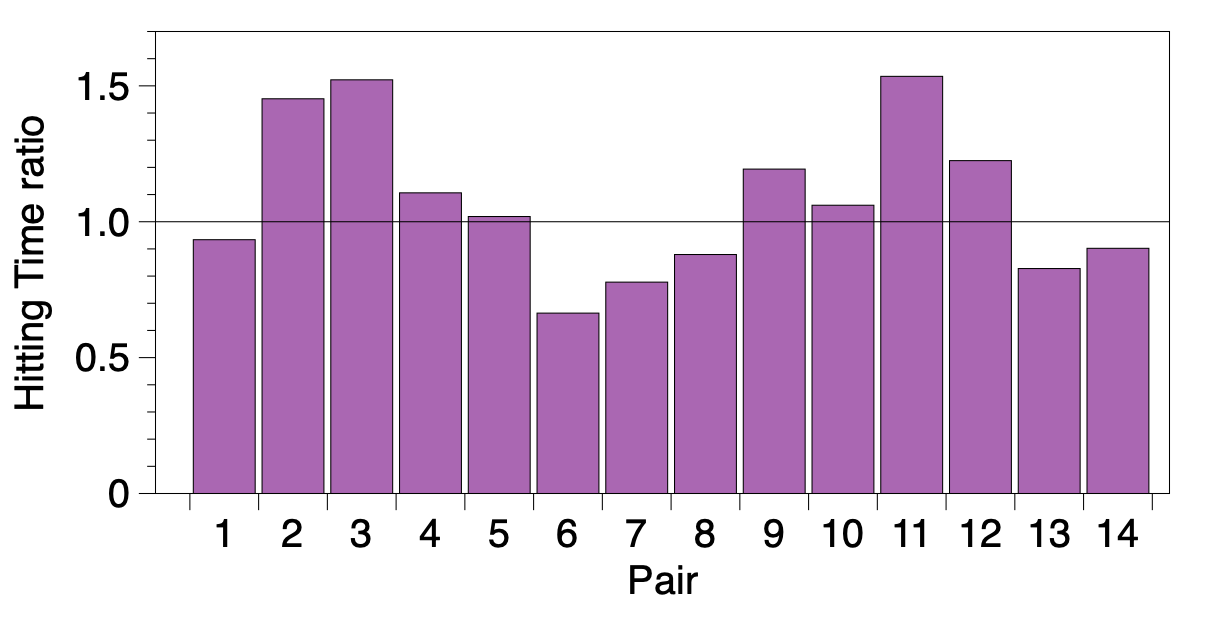}
\par\end{centering}
\caption{{\small{Excess Time Saving (left) and Hitting Time Ratio (right, see the text for details) for 14 out of 20 pairs of nodes in the brain network, for which there exist a SP and a SRP where SP-type and SRP-type walkers are found. Results are in agreement with the idea particles can get lost easier if random-walking through the SP than through the SRP. }}}
\label{fig:ETS_HTR}
\end{figure}

{As explained in the paper, a free-flowing particle hopping randomly throughout the network is more likely to spread out and get lost through local connectivity paths (through the paths arising in intermediate nodes) if the path under study is proportionally more resistive.\\
Using random walks between a source node $i$ and a destination node $j$ can allow us to quantify the excess time needed to traverse different types of paths connecting $i$ and $j$, and thus, in principle we can use random walk dynamics to illustrate  the effect that bypasses have on network navigability by random walkers.}
\paragraph{{ S6.1 Analysis in the toy network}}
{
Since the toy network in Fig.\ref{fig:cartoon} is small, we can exhaustively analyse the statistics of random walkers. In particular, we have generated $10^5$ random walkers starting at $i=p$ and monitor their node sequence until each of them ends at $j=q$, that is, we record the full sequence of nodes of a walker along its walk from $p$ to eventually reaching $q$. Subsequently, each walker is then {\it classified} as a SRP-type or a SP-type depending on the sequence of nodes they travel (this task is easy to do in the toy network, as for instance walkers travelling the SRP reach the destination from a node which is different from the one used if traversing the network via a SP region, see Fig.\ref{fig:cartoon}). Once walkers have been classified, we then proceed to compare both classes by computing two magnitudes that proxy the time needed for the walker to reach destination and play the role of a dynamic (random-walk-like) analog of our network geometrization:}
{\begin{itemize}
\item The hitting time $HT$, defined as the total time required on average (average over all walkers in the same class) to reach the destination node $q$. This time is simply computed as the average size of the node sequences. It is a dynamic proxy for $\mathbb{L}$. To compare hitting times for both classes, we also compute the Hitting Time Ratio 
$$HRT=\frac{HT_{\text{SRP-type}}}{HT_{\text{SP-type}}},$$
where $HRT<1$ indicates that the net time spent by SRP-type walkers is lower than SP-type walkers. 
\item The Excess Time $ET$, that measures how much more time a random walker using e.g. the bypass needs due to the fact that it can backtrack or get lost in the vicinity of the bypass as compared to the time it would take if the walker was a ballistic one, i.e.
$$ET_{\text{SRP-type path}}  = \frac{\langle HT_{\text{SRP-type}} \rangle} {\text{number of nodes in  the SRP}}$$
$$ET_{\text{SP-type path}} = \frac{\langle HT_{\text{SP-type}} \rangle}{\text{number of nodes in  the SP}}$$
Note that the excess time $ET$ can be seen as a random-walk dynamics equivalent of 
the magnitude $\mathbb{L}/\ell$, and thus in the same vein we have that $ET_{\text{SRP-type path}}/ET_{\text{SP-type path}}$ is the random-walk dynamics equivalent of the measure $\rho$ (Eq.\ref{eq:rho}).
\item The Excess Time Saving $ETS$ is a percentage that compares the Excess Time via SRP and SP:
$$ETS=\frac{ET_{\text{SP-type path}}-ET_{\text{SRP-type path}}}{ET_{\text{SRP-type path}}}\times 100$$
\end{itemize}}
{For the toy network, we find 
$\langle HT_{\text{SRP-type}}\rangle \approx 6.9$, whereas $\langle HT_{\text{SP-type}}\rangle \approx 10.9$, i.e. the hitting time is smaller for walkers of the SRP-type class  than for the SP-type class. Similarly, we find  $ET_{\text{SRP-type path}}\approx 1.38, ET_{\text{SP-type path}}\approx 2.7$, i.e. the time lost by SRP-type walkers due to their stochasticity is half of the time lost by SP-type walkers. }
\paragraph{{S6.2 Analysis in a real network}}
{We extend the previous analysis to a (larger) real network (functional --task-driven-- brain network, see SI section on Empirical networks for details), composed by $n=638$ nodes.
We run $2\times 10^4$ random walks between source $i$ and destination $j$, for a total of $M=20$ pairs of nodes $(i,j)$ where we identified the SP and the SRP.
Each of the walkers starts at $i$, makes a random excursion on the network, and eventually reaches $j$. Each walker is then {\it classified} as a SRP-type if all the nodes of the SRP are visited by the walker and none of the nodes of the SP is visited. Otherwise, the walker is classified as a SP-type if all the nodes in the SP are visited, but not all the SRP nodes are visited. The rest of the walkers are labelled as hybrid and not considered in the analysis. Finally, we compute the average hitting time and the excess time for each class and pair $(i,j)$, and the excess time saving for each pair $(i,j)$. Our results, depicted in Fig.\ref{fig:ETS_HTR}, indicate that the excess time of SP-like walkers is systematically larger than the excess time of SRP-like walkers, on agreement with previous results on the toy network. We find mixed results for the hitting times: for some pairs the average time spent is higher in the SP-type class, for some pairs it is higher in the SRP-type class. Finally, the excess time saving is systematically larger than 0, fluctuating between $25\%$ and $225\%$. All in all, these results --while not exhaustive by any means-- support and reinforce the idea that particles can get lost easier if random-walking through the SP than through the SRP.}

\subsection*{S7 Further effects on dynamics: synchronization and epidemic spreading}

\begin{figure}[t!]
\begin{centering}
\includegraphics[width=0.83\textwidth]{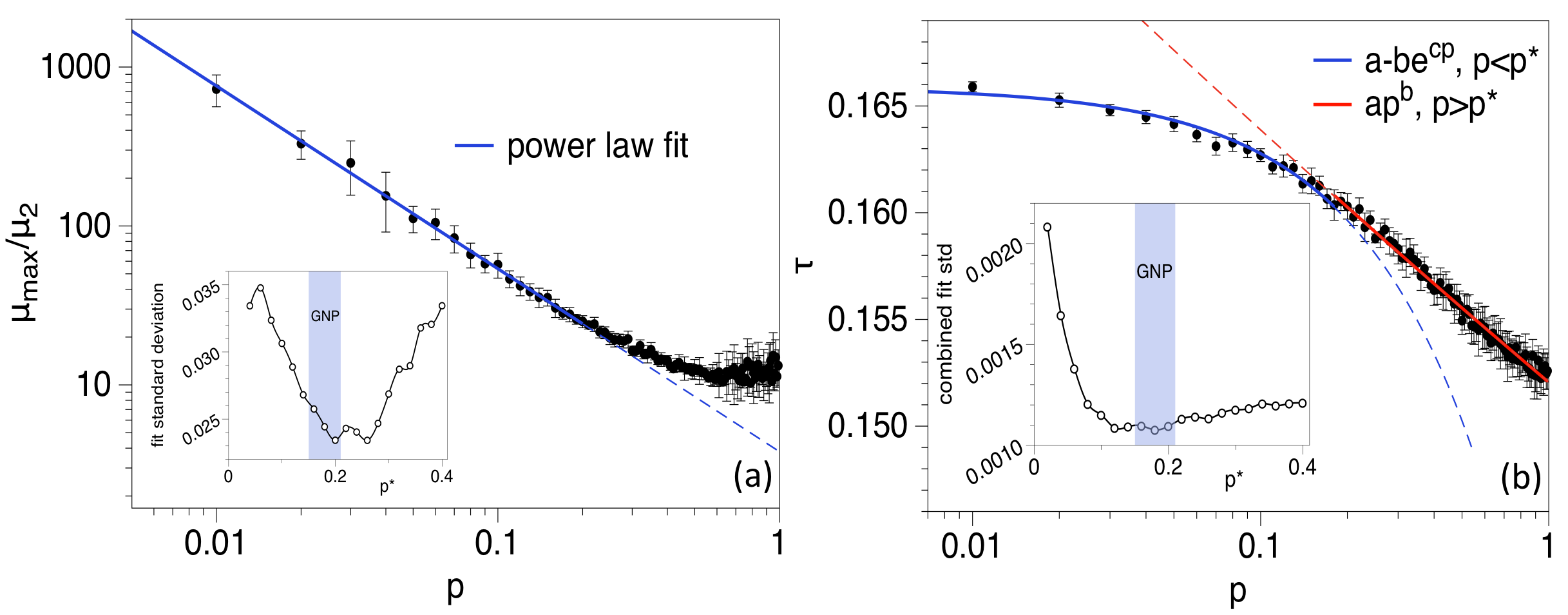}
\par\end{centering}
\caption{{\small{\bf Dynamics.} (a) Plot of the eigenratio $\mu_{\textnormal{max}}/\mu_{2}$ of the Laplacian matrix versus the rewiring probability $p$ in a WS model with $N=500$ nodes and average degree $k=6.$ The straight line corresponds to the power law decay of the eigenratio $\sim p^{-1.15}$. Note that the power law decay is a very good fit only in the range $0<p<p^*$, thereafter smoothly crossovering to a flat dependence. (Inset) Sample standard deviation of the power law fit as a function of the upper bound $p^*$. This fitting error is minimized close to $p^*\approx p_{\text{GNP}}$, i.e. the crossover between a power law decay of the eigenratio and a flat dependence on $p$ takes place at the network's Good Navigational Point, where bypasses have a maximal effect. (b) Plot of the epidemic threshold $\tau=1/\lambda_{1}$ as a function of the rewiring probability $p$ in the WS model with $N=500$ nodes and average degree $k=6$ (dots and error bars denote average and standard deviation over 100 realizations). The blue and red lines correspond to two qualitatively different fits, for $p<p^*$ (blue) and $p>p^*$ (red). 
(Inset) To find the value of $p^*$ that marks the crossover between scaling behaviors, we compute the aggregate fit deviation of the two-function model, finding that such deviation reaches a minimum for $p^*\approx p_{\text{GNP}}$, i.e. the change of behavior onsets at the regime where bypasses take a predominant role.}
\label{fig:dynamics}}
\end{figure}

\paragraph{S7.1 Synchronization} The synchronizability of a network of coupled oscillators is governed
by the Laplacian eigenratio $\mu_{\textnormal{max}}/\mu_{2}$,  where $\mu_{\textnormal{max}}$ and $\mu_{2}$ are the largest and smallest (non-null) eigenvalues of the network's Laplacian matrix, such that the smaller such eigenratio, the more synchronizable the network is.
In panel (a) of Fig. \ref{fig:dynamics}
we illustrate the change of such eigenratio with the rewiring probability $p$
in the Watts-Strogatz model. Interestingly, we find a qualitative change of behavior at $p^*\approx p_{\text{GNP}}$, where for $p<p^*$ the eigenratio scales as  $\mu_{\textnormal{max}}/\mu_{2} \sim p^{-\gamma}$, with $\gamma\approx 1.153$,  whereas for $p>p^*$ the eigenratio's behavior smoothly crossovers to a flat dependence. To derive the crossover value, we compute the sample standard deviation 
\begin{equation}
s=\sqrt{\sum_{i=1}^{n(p^*)}(y_i - f(x_i))^2/(n(p^*)-1)}
\label{eq:s}
\end{equation}
between the actual eigenratio and the scaling function above in the range $p<p^*$ and find the value $p^*$ which minimises such error (see inset of panel (a)  in Fig. \ref{fig:dynamics}). We find that the error in the fit is minimised for $p^*\approx 0.2$, indeed close to $p_{\text{GNP}}$.

\paragraph{S7.2 Epidemic spreading} 
The second process is a SIR-type epidemic spreading over the network, where $\alpha$ is the birth rate of the epidemic and $\delta$
is its death rate. It is well known that when $\alpha/\delta<\tau$
infection dies out and when $\alpha/\delta>\tau$ infection survives
and becomes an epidemic, where $\tau$ is the epidemic threshold (the smallest $\tau$, the easiest that an infection outbreak becomes epidemic). For undirected networks $\tau=1/\lambda_1$, where $\lambda_1$
is the spectral radius of the adjacency matrix. 
As can
be seen in panel (b) of Fig. \ref{fig:dynamics} the epidemic threshold of a Watts-Strogatz
network also shows a qualitative change: there is a crossover from a slow decrease for $p<p^*$ to a faster, power-law decrease for $p>p^*$. To find $p^*$, we compute the sample standard deviation of each branch (eq.\ref{eq:s}). The inset of panel (b) shows the combined sample standard deviations, that find a minimum for $p^*\approx p_{\text{GNP}}$.
That is,
up to the point in which bypasses become predominant, an infection
is relatively easier to control than after that point.\\
Altogether, these results suggest that dynamical processes running on the network perceive such bypasses in a nontrivial way, and these therefore play a role in the network's function.

\subsection*{S8 Additional details on the $\epsilon$ vs $\langle k \rangle$ in ER and SW networks}

\begin{figure}[t!]
\begin{centering}
\includegraphics[width=0.4\textwidth]{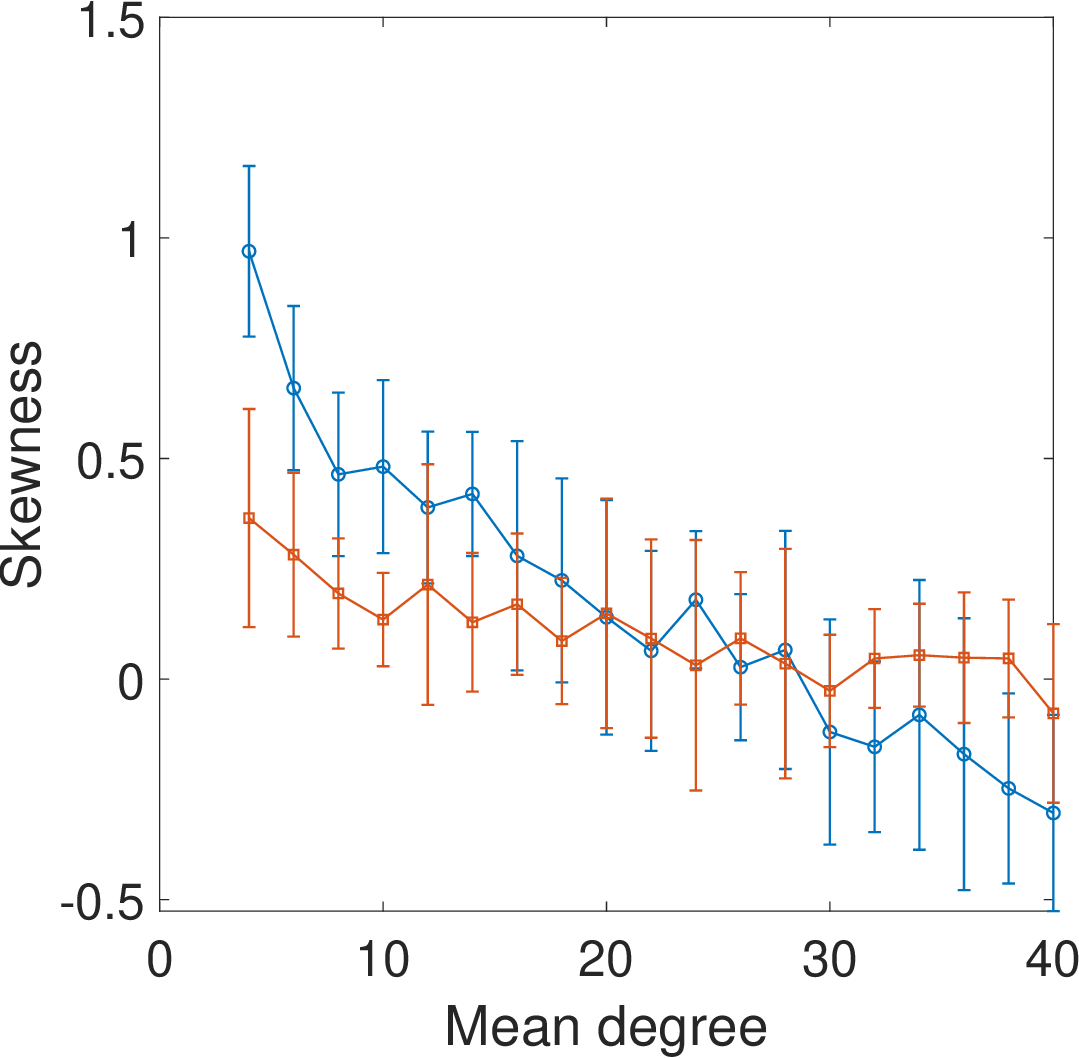}
\includegraphics[width=0.38\textwidth]{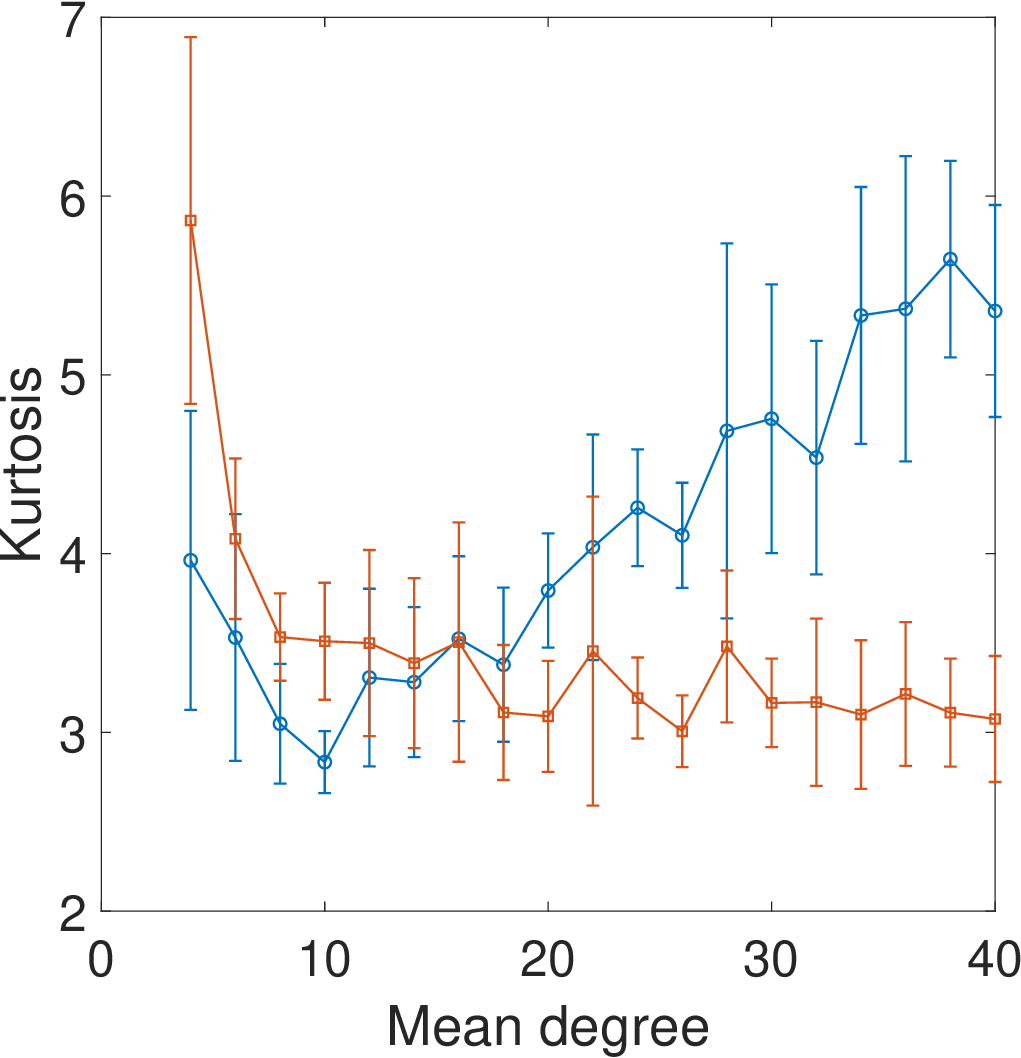}
\par\end{centering}
\caption{{\small Plots of skewness (left) and kurtosis (right) of the degree distribution of ER networks and GNP-SW networks (with $p=p_{\text{GNP}}=0.15$) as a function of their mean degree $\langle k \rangle$.}
\label{fig:skew_kurt}}
\end{figure}

\begin{figure}[t!]
\begin{centering}
\includegraphics[width=0.85\textwidth]{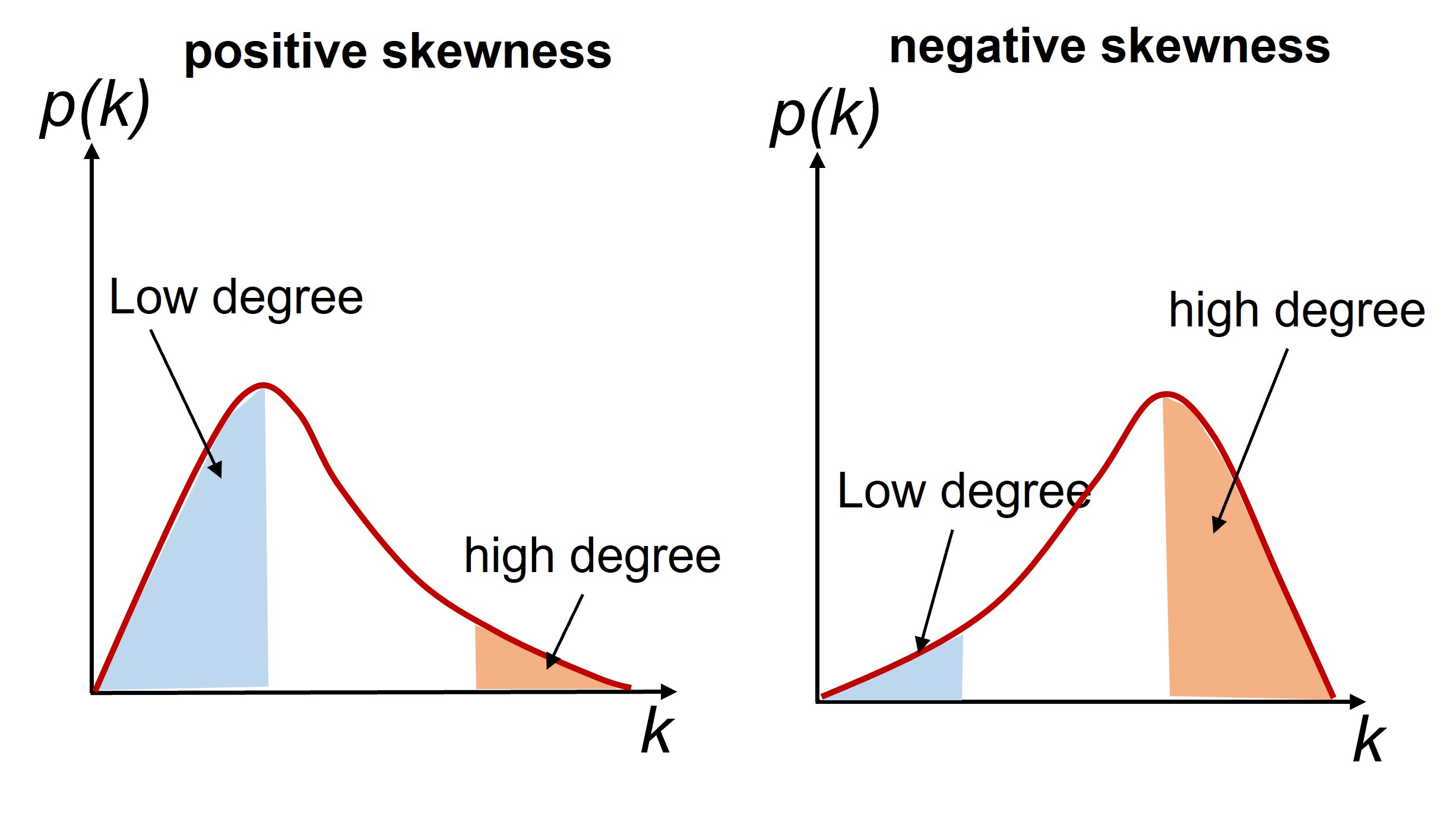}
\par\end{centering}
\caption{{\small A sketch of the degree distribution with positive and negative skewness respectively, and the relative abundance of low and high degree nodes.}
\label{SK}}
\end{figure}

Let us look at the degree distributions of the WS networks with $p=0.15$ generated for very different $\left\langle k\right\rangle $. When $\left\langle k\right\rangle $ is relatively low the WS generates networks with a significant ``tailedness'' of the degree distribution. This is because a node with relatively low degree to which a small number of edges are removed from or added to, changes significantly its degree relative to the original one, e.g., a node of degree two to which tow edges are added duplicates its degree relative to the original one. This effect is diluted, however, as soon as $\left\langle k\right\rangle $ increases. Therefore, for relatively small $\left\langle k\right\rangle $ we should find the existence of certain outlier nodes in terms of their degrees. This is exactly what it is observed when we consider the kurtosis of the
degree distributions of WS networks (see Fig.\ref{fig:skew_kurt}). For relatively small $\left\langle k\right\rangle $ the kurtosis of the degrees in SW
networks is bigger than that of ER networks. When $\left\langle k\right\rangle >16$ this relation is inverted and ER displays more ``tailedness'' than the WS degrees.\\
But this is not the end of the story. The asymmetry of the degree distribution of an ER network also changes dramatically when $\left\langle k\right\rangle $ increases. For $\left\langle k\right\rangle \leq30$
the skewness of the degree in ER networks is positive, and indeed bigger than that of a WS network of comparable size. This means the existence of significantly more low degree nodes than hubs, i.e., it looks like a mini-power-law. However, for $\left\langle k\right\rangle >30$ the skewness of the degrees in ER become negative and smaller than that of WS, which remains positive (see Fig.\ref{SK} for a sketch). This gives again advantages to WS networks to avoid these few hubs existing in it and using bypasses. However, the ER now have much more high degree nodes than low degree ones, making impossible to have bypasses. Does a GNP exists for this regime, $\left\langle k\right\rangle >30$,
and if so is it at $p=0.15$? The answer to both questions is positive.

\subsection*{S9 Super-hubs and the transition to ultra-short graphs}

\begin{figure}[t!]
\begin{centering}
\includegraphics[width=0.8\textwidth]{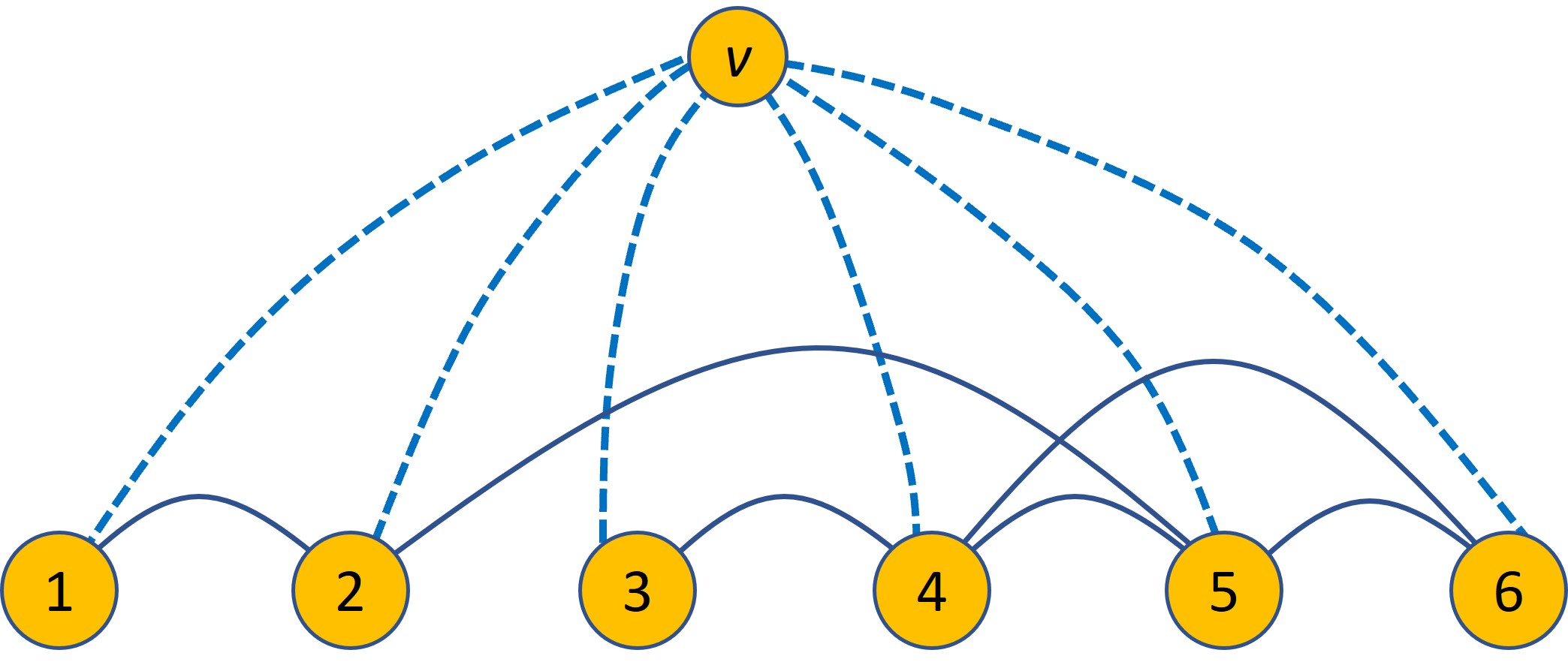}
\par\end{centering}
\caption{\small{Illustration of the umbrella effect. Initially we have a simple sparse connected network
$G$ with 6 nodes and 6 edges.  A new graph $G'$ is created by
adding a node connected to every existing node of $G$. This new super-hub makes all previous paths in the original network to be bypasses of the shortest paths, that now involve the super-hub. Adding this super-hub makes $\epsilon$ to experience a sudden increase, what we call the umbrella effect.}}
\label{fig:umbrella}
\end{figure}

As we have seen the preferential attachment method produces significantly
more bypasses than a random interconnection of nodes. The question
that emerges is if there is a simple mechanism which produces even
more bypasses than those introduced by the skew degree distribution
of the BA method. Let us consider a sparse connected random graph
$G$ (see Fig. \ref{fig:umbrella}). The chances that there are some
bypasses in $G$ are scarce, so we can consider that such number is
nill. As the graph $G$ is connected there is a SP between every pair
of nodes. Now, let us create $G'$ by adding a new node connected
to every node of $G$. This new node
$v$ is a super-hub, which is connected to $n'-1$ nodes, where $n'$
is the number of nodes in $G'.$ Therefore, a particle trying to go
from one node $i\neq v$ to another $j\neq v$ will avoid going through
the super-hub as this will be a very resistive trajectory. However,
the particle can travel by using the SP connecting them in $G$. For
instance, it can go through the path: $1-2-5$, $1-2-5-6$, $1-2-5-4$,
$1-2-5-4-3$ instead of through the paths $1-2-j$. Consequently,
all those paths which are of length larger than 2 become potential
bypasses by the effect of the newly added super-hub. Due to the shape
of the graph illustrated in Fig. \ref{fig:umbrella} we call this the
``umbrella effect'', and relates to the transition to ultra-short graphs. In closing, this mechanism potentially transforms
all SP of length larger than 2 in a sparse graph into bypasses.\\
This effect is expected to occurs in the BA mechanism when
the mean degree $k_{mean}$ of the network is large. That
is, suppose that at certain ``time'' $t$ of the BA evolutionary
mechanism, the number of existing nodes $n_{t}$ is smaller than $\langle k \rangle$.
This frequently happens at the initial stages of the evolution of
BA networks when $\langle k \rangle$ is relatively large. In these cases,
the new nodes attached to the network need to be connected to every
existing node in the graph, producing an umbrella effect during these
periods of the process. The resulting effect could be an explosion
of bypasses in the final BA network when $\langle k \rangle$ is relatively
large. 

\subsection*{S10 Discussion on weighted networks}

{When we consider a generalized communicability distance $\xi_{ij}\left(\beta\right)=\sqrt{G_{ii}\left(\beta\right)+G_{jj}\left(\beta\right)-2G_{ij}\left(\beta\right)}$,
the values of the three terms involved in $\xi_{ij}\left(\beta\right)$
change nonlinearly with $\beta$. Consequently, there will be effects
on the lengths of the paths studied which depends on this parameter.
To illustrate this situation let us consider our toy network in which
we have found that $\epsilon=0.6944$ due to the existence of the
bypass 1-9-8-7-4. If we multiply the adjacency matrix by 0.5, i.e.,
we change the weight of every edge from 1.0 to 0.5. In this case we
obtain $\epsilon=0$, which means that the bypass no longer exists.
This means that if we reduce the weight of every edge by a half, it
is no longer ``economic'' for a walker to use the path 1-9-8-7-4
instead of the path 1-2-3-4. The fact is that when $\beta=1$ the
communicability distance between 2 and 3 is slightly longer than that
between 8 and 9. However, when $\beta=0.5$ both distances are identical,
such that it is shorter (in terms of communicability distances) going
through the topological SP. We use this example to find the meaning
of the edge weights in the context of communicability distances. For
that we can use the metaphor that the edge weight represents the ``width''
of the edge like if it were a street connecting two intersections.
The influence of this width is twofold. For instance, suppose that
two nodes $i$ and $j$ are connected by an edge with weight equal
to 2. This means that we have doubled the width of the street by adopting
two lanes to connect the pair of nodes. It is evident that this will
increase the communicability $G_{ij}$ between the two nodes as well
as of those in any path containing that edge. However, the ``complexity''
of the intersctions $i$ and $j$ is also increased, which means an
increase in the values of $G_{ii}$ and $G_{jj}.$ The distance $\xi_{ij}$
will depend on how much each of these terms, $G_{ij}$ or $G_{ii}$
and $G_{jj}$, increases more. A similar situation emerges if we consider
that the edge weight is for instance $0.1$, which indicates that
we have dropped the width of the corresponding street. This drop in
the edge weight produces a decrease in the three terms implied in
the definition of the communicability distance. Consequently, it will
depend on the topological characteristics of the nodes and edge involved
whether the distance decrease significantly of not in relation to
the unweighted case.\\}

{In order to be able to compare a weighted and an unweighted network
we need to normalize the weights of a network such as their mean is
unity coinciding with the mean weight of the edges in a non weighted
network. This is what we do in the next examples considering weighted
networks. Let us now provide some numerical examples based on our
toy model. First, let us consider the edge $8-9$ which is in the
SCP but not in the SP. If we increase the value of this weight, for
instance to values between $1.5$ to $2.0$, we obtain that $\epsilon=0$,
indicating that there is no longer a bypass between the nodes $1$
and $4$. In this case any particle traveling between the nodes 1
and 4 will prefer to go through the SP. This result indicates that
here the increase in the self-communicabilities of the nodes 8 and
9 is more significant than the increase of the communicability between
the two nodes. As a consequence the communicability distance between
the nodes 8 and 9 increases significantly making the traveling through
them more difficult than through the SP. If we drop the value of this
weight between 0.1 and 1.4, the increase of the communicability distance
between 8 and 9 is not enough as to make the traveling more efficient
through the SP. Consequently, we obtain $\epsilon=0.6944$, indicating
the existence of a bypass.\\} 

\begin{figure}[htb]
\includegraphics[width=0.6\textwidth]{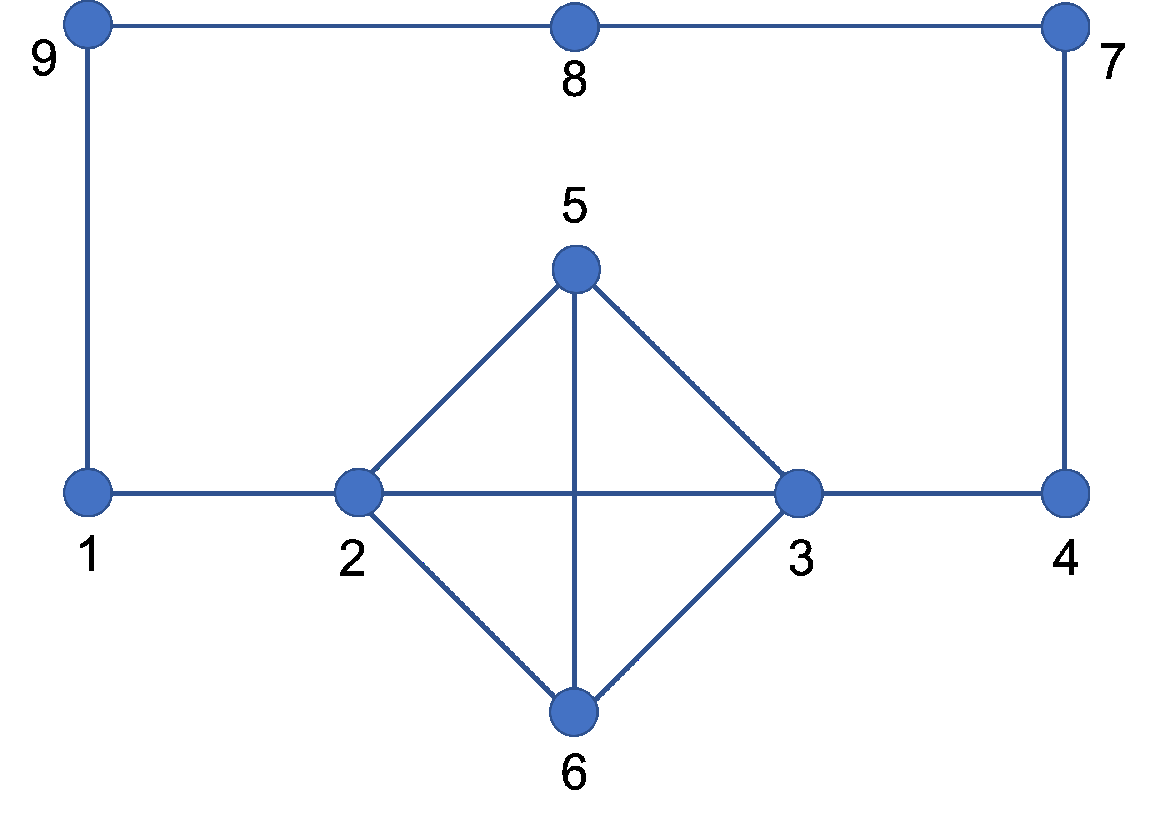}
\caption{Toy network with node labelling.}
\end{figure}

{To illustrate the complexities of this problem let us increase the
weight of this edge too much as to facilitate even more the navigation
through the SP connecting the nodes 1 and 4. For instance, if we increase
this weight up to 10, it is true that a walker travelling between
1 and 4 will go through the SP connecting them. However, the result
of the whole network is that $\epsilon=0.6944$. This indicates that
another bypass has emerged. This occurs for a particle traveling between
1 and 7, for which it is preferable to go through the path 1-2-3-4-7
than through the path 1-9-8-7, because of the large weight of an edge
in the second path.\\
To continue illustrating the nontriviality of the influence of edge
weights on the communicability distance let us select the weight between
the nodes 2 and 3. This edge is in the SP between nodes 1 and 4 and
not in the SCP. Dropping this weight will facilitate the navigation
through the SP. The question is whether this effect is enough as for
the particle to prefer to go through the path 1-2-3-4 than through
the bypass 1-9-8-7-4. The answer is not. If we drop this weight to
0.1 we find $\epsilon=0.6944$ and the bypass between the two nodes
remains. Not even dropping it to $10^{-12}$ makes any difference.
The question is that the node 2 is still connected to three other
nodes with a weight equal to one, which still keep a large weighted
self-communicability for this node due to its implication in triangles,
squares, etc. To make that the bypass disappears we need to change
the weights of the 6 edges in the central clique of the graph. In
this case, if for instance, we drop their weights to 0.1, the bypass
no longer exists and the particle will travel through the SP. What
happened here is that by changing the weights of all edges in the
clique we have reduced significantly the values of the self-communicabilities
of the nodes 2 and 3 as to make that the communicability distance
drops significantly enough as to facilitate the navigation through
this path.}

\subsection*{S11  Empirical networks}

\paragraph{S11.1 Additional details on empirical networks}
Below we provide details on all empirical networks, grouped by topic (see also table \ref{table:nets} for a summary of some characteristics).\\

\noindent \textit{Brain networks }
\begin{itemize}
\item {\bf Neurons:} Neuronal synaptic network of the nematode \textit{C. elegans}.
Included all data except muscle cells and using all synaptic connections
\cite{Milo datasets}
\item {\bf Human brain (anatomical) \cite{brain}}: Structural
connectivity network of 70 healthy volunteers studied by Hagmann et
al. The nodes of these networks correspond to 1,000 physical regions
in the brain and the connections among them were estimated for individual
participants using deterministic streamline tractography, such that
they represent physical connections between brain areas.
\item {\bf Human brain (functional, resting-state) \cite{brain}}: We then studied the functional brain networks for the same 70
individuals studied while participants were not engaged in any overt
task, such that the scans are treated as eyes-open resting-state fMRI
among the same 1,000 brain regions studied before. In this case we
used a threshold for each brain network such that the resulting webs
had the same edge densities than the corresponding structural connectivity
ones. 
\item {\bf Human brain (functional, task-driven)}: task-related coactivation brain network Although this
network has a different parcellation that the previous two ones (consisting
on 638 brain regions), this is the product of a meta-analysis of the
large primary literature that used fMRI or PET to measure task-related
activation (meta-analysis of over 1,600 studies published
between 1985--2010 \cite{Brain networks datasets}).
\end{itemize}
\noindent \textit{Ecological networks }
\begin{itemize}
\item {\bf Benguela}: Marine ecosystem of Benguela off the southwest coast of
South Africa \cite{Benguela}
\item {\bf Bridge Brook}: Pelagic species from
the largest of a set of 50 New York Adirondack lake food webs \cite{Bridge Brooks};
\item {\bf Canton Creek}: Primarily invertebrates and algae in a tributary, surrounded
by pasture, of the Taieri River in the South Island of New Zealand
\cite{Canton}; 
\item {\bf Chesapeake Bay}: The pelagic portion of an eastern
U.S. estuary, with an emphasis on larger fishes \cite{Chesapeake};
\item {\bf Coachella}: Wide range of highly aggregated taxa from the Coachella
Valley desert in southern California \cite{Coachella}; \item {\bf El Verde}:
Insects, spiders, birds, reptiles and amphibians in a rainforest in
Puerto Rico \cite{ElVerde}; \item {\bf Grassland}: all vascular plants and all
insects and trophic interactions found inside stems of plants collected
from 24 sites distributed within England and Wales \cite{Grassland};
\item {\bf Little Rock}: Pelagic and benthic species, particularly fishes, zooplankton,
macroinvertebrates, and algae of the Little Rock Lake, Wisconsin,
U.S. \cite{LittleRock}; \item {\bf Reef Small}: Caribbean coral reef ecosystem
from the Puerto Rico-Virgin Island shelf complex \cite{Reef Small};
\item {\bf Scotch Broom}: Trophic interactions between the herbivores, parasitoids,
predators and pathogens associated with broom, Cytisus scoparius,
collected in Silwood Park, Berkshire, England, UK \cite{Scotch Broom};
\item {\bf Shelf}: Marine ecosystem on the northeast US shelf \cite{Shelf}; \item {\bf Skipwith}:
Invertebrates in an English pond \cite{Skipwith}; \item {\bf St. Marks}: Mostly
macroinvertebrates, fishes, and birds associated with an estuarine
seagrass community, Halodule wrightii, at St. Marks Refuge in Florida
\cite{StMarks};\item {\bf St. Martin}: Birds and predators and arthropod prey
of Anolis lizards on the island of St. Martin, which is located in
the northern Lesser Antilles \cite{StMartin}; \item {\bf Stony Stream}: Primarily
invertebrates and algae in a tributary, surrounded by pasture, of
the Taieri River in the South Island of New Zealand in native tussock
habitat \cite{Stony}; \item {\bf Ythan1}: Mostly birds, fishes, invertebrates,
and metazoan parasites in a Scottish Estuary \cite{Ythan1} ; \item {\bf Ythan2}:
Reduced version of Ythan1 with no parasites \cite{Ythan2}. 
\item {\bf Termite}: The networks of three-dimensional galleries in termite nests
\cite{Termite Moulds}; 
\end{itemize}
\newpage
\noindent \textit{Informational networks }
\begin{itemize}
\item {\bf Roget}: Vocabulary network of words related by their definitions in
Roget\textquoteright s Thesaurus of English. Two words are connected
if one is used in the definition of the other \cite{Roget}; 
\end{itemize}
\textit{Biological networks }
\begin{itemize}
\item {\bf Protein-protein interaction networks} in: \textit{S. cerevisiae} (yeast)
\cite{PIN_yeast_1,PIN_yeast_2}; 
\item {\bf TransE.coli, Transyeast}: Direct transcriptional regulation between
genes in \textit{Saccaromyces cerevisae}. \cite{Milo datasets,MiloDatasets2}. 
\end{itemize}
\textit{Social networks }
\begin{itemize}
\item {\bf Geom}: Collaboration network of scientists in the field of Computational
Geometry \cite{Batagelj Mrvar}; 
\item {\bf QcGr} Collaboration network of scientists
in the field of Quantum Gravity \cite{GrQc}; 
\item {\bf Drugs}: Social network
of injecting drug users (IDUs) that have shared a needle in the last
six months \cite{Moody}; 
\end{itemize}
\textit{Technological and infrastructural networks }
\begin{itemize}
\item {\bf Electronic}: Three electronic sequential logic circuits parsed from
the ISCAS89 benchmark set, where nodes represent logic gates and flip-flop
\cite{Milo datasets}; 
\item {\bf USAir97}: Airport transportation network between
airports in US in 1997 \cite{Batagelj Mrvar}; 
\item {\bf Internet}: The internet
at the Autonomous System (AS) level as of September 1997  \cite{Internet};
\item {\bf Power Grid}: The power grid network of the Western USA \cite{WattsStrogatz}. 
\end{itemize}
\newpage
\textit{Software networks }
\begin{itemize}
\item Collaboration networks associated with six different {\bf open-source software}
systems, which include collaboration graphs for three Object Oriented
systems written in C++, and call graphs for three procedural systems
written in C. The class collaboration graphs are from version 4.0
of the {\bf VTK} visualization library; the CVS snapshot dated 4/3/2002
of Digital Material (DM), a library for atomistic simulation of materials;
and version 1.0.2 of the AbiWord word processing program. The call
graphs are from version 3.23.32 of the {\bf MySQL} relational database system,
and version 1.2.7 of the {\bf XMMS} multimedia system. Details of the construction
and/or origin of these networks are provided in Myers \cite{Myers Software}.
\end{itemize}

\begin{table}[t!]
\begin{centering}
\begin{tabular}{|c|c|c|}
\hline 
{\bf Name} & $n$ & $m$\tabularnewline
\hline 
Benguela & 29 & 191\tabularnewline
BridgeBrook & 75 & 542\tabularnewline
Canton & 108 & 707\tabularnewline
Chesapeake & 33 & 71\tabularnewline
Coachella & 30 & 241\tabularnewline
Drugs & 616 & 2012\tabularnewline
Electronic1 & 122 & 189\tabularnewline
Electronic2 & 252 & 399\tabularnewline
Electronic3 & 512 & 819\tabularnewline
ElVerde & 156 & 1439\tabularnewline
Collaboration CoGe & 3621 & 9461\tabularnewline
Collaboration QcGr & 4158 & 13428\tabularnewline
Internet1997 & 3015 & 5156\tabularnewline
LittleRockA & 181 & 2318\tabularnewline
Neurons & 280 & 1973\tabularnewline
PIN\_Yeast & 2224 & 6608\tabularnewline
Power\_grid & 4941 & 6594\tabularnewline
ReefSmall & 50 & 503\tabularnewline
Roget & 994 & 3640\tabularnewline
ScotchBroom & 154 & 366\tabularnewline
Shelf & 81 & 1451\tabularnewline
Skipwith & 35 & 353\tabularnewline
Software\_Abi & 1035 & 1719\tabularnewline
Software\_Digital & 150 & 198\tabularnewline
Software\_Mysql & 1480 & 4140\tabularnewline
Software\_VTK & 771 & 1357\tabularnewline
Software\_XMMS & 971 & 1802\tabularnewline
StMarks & 48 & 218\tabularnewline
StMartin & 44 & 218\tabularnewline
Stony & 112 & 830\tabularnewline
Termite\_1 & 507 & 676\tabularnewline
Termite\_2 & 260 & 280\tabularnewline
Termite\_3 & 268 & 437\tabularnewline
Transc\_yeast & 662 & 1062\tabularnewline
USAir97 & 332 & 2126\tabularnewline
Ythan1 & 134 & 593\tabularnewline
Ythan2 & 92 & 416\tabularnewline
Brain (task-driven) & 638 & 18625\tabularnewline
Brain (resting-state) & 1000 & $10724 \pm 708$\tabularnewline
\hline 
\end{tabular}
\par\end{centering}
\caption{Dataset of real-world networks studied in this paper,
their size $n$ (number of nodes), and number of edges $m$.}
\label{table:nets}
\end{table}

\begin{figure}[t!]
\begin{centering}
\includegraphics[width=0.7\textwidth]{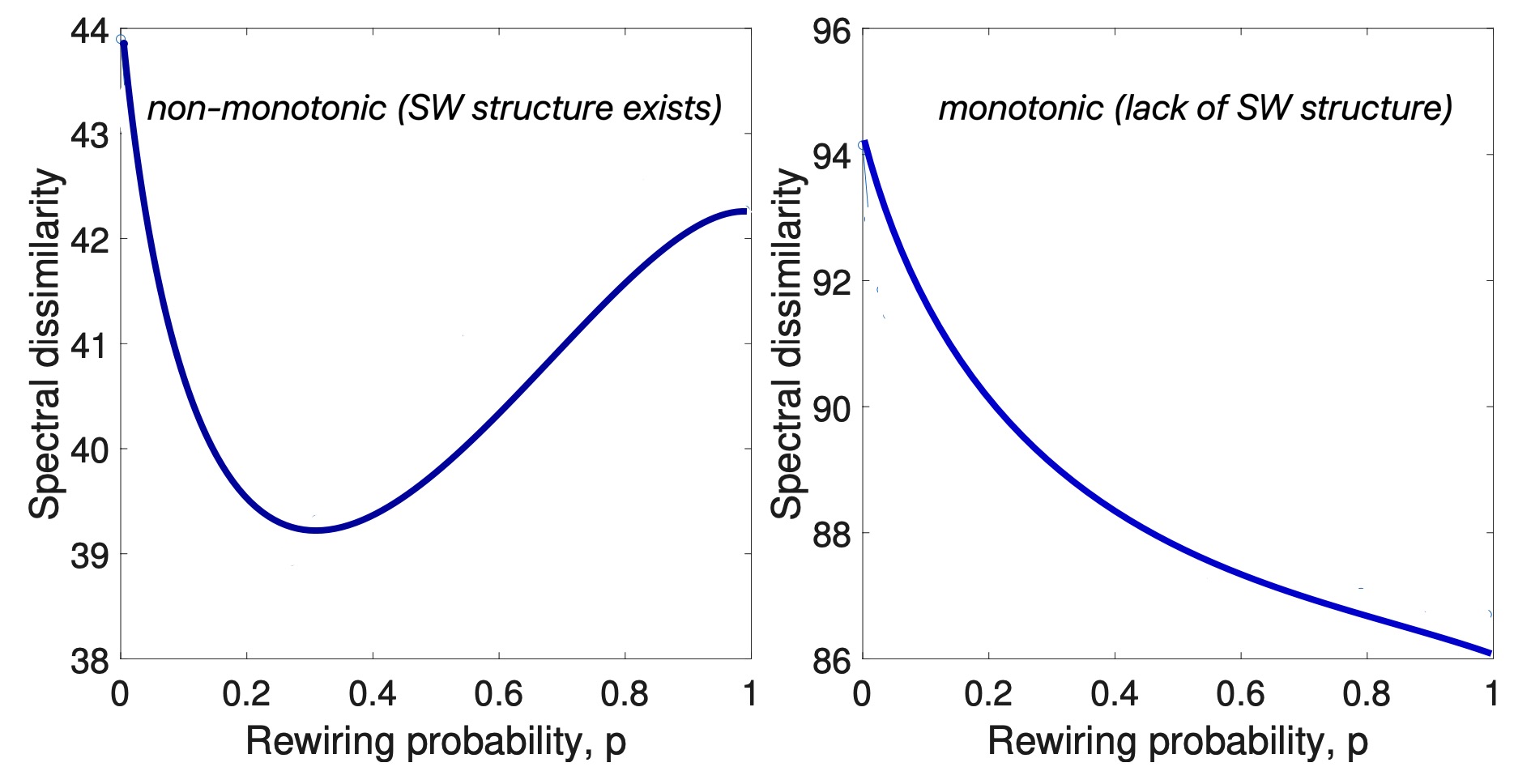}
\par\end{centering}
\caption{Typical dissimilarity plots obtained in the analysis of empirical networks: we either we find a non-monotonic behavior that highlights an optimal value at intermediate values of $p$ (left panel) or a monotonically decreasing function, for which $p^*=1$ (right panel). The former suggests a clear SW structure while the latter suggests that the network is closer to a random graph.}
\label{similarity}
\end{figure}

\begin{figure}[t!]
\begin{centering}
\includegraphics[width=1.0\textwidth]{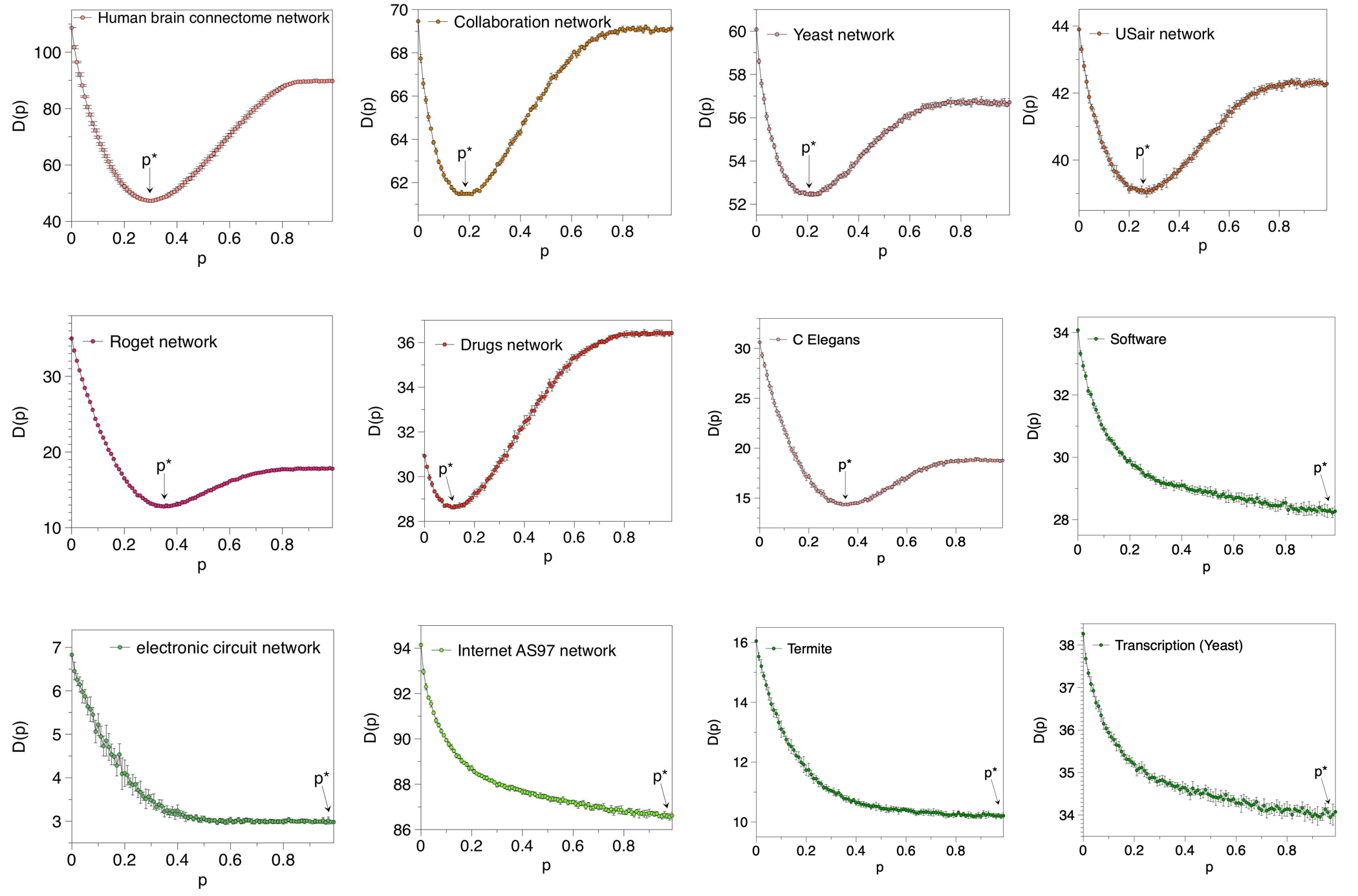}
\par\end{centering}
\caption{Dissimilarity curves ${\cal D}(p)$ as a function of the rewiring probability $p$, for different empirical networks: biological (human brain, C-elegans, protein-protein interaction, transcription, termite mound), social (drugs, scientific collaboration, Roget thesaurus), technological (airports, Internet AS, software, electronic circuits).  The minimum of these curves characterizes the type of SW network which is more similar to the empirical network.}
\label{similarity_all}
\end{figure}

\paragraph{S11.2 Additional details on calculation of $p^*$}
As can be seen in  table 1 of the main article, 
we find two classes of networks, one class is formed by networks
for which a minimum dissimilarity exists between the real-world network
and a WS model operating close to $p_{GNP}$ (see Fig. \ref{similarity}, left panel). The other class is formed
by networks 
which are more similar to the random graph, i.e., WS with $p=1$  (see Fig. \ref{similarity} right panel). The dissimilarity curve is plotted for some empirical networks in Fig. \ref{similarity_all}.

\begin{figure}[t!]
\begin{centering}
\includegraphics[width=0.85\textwidth]{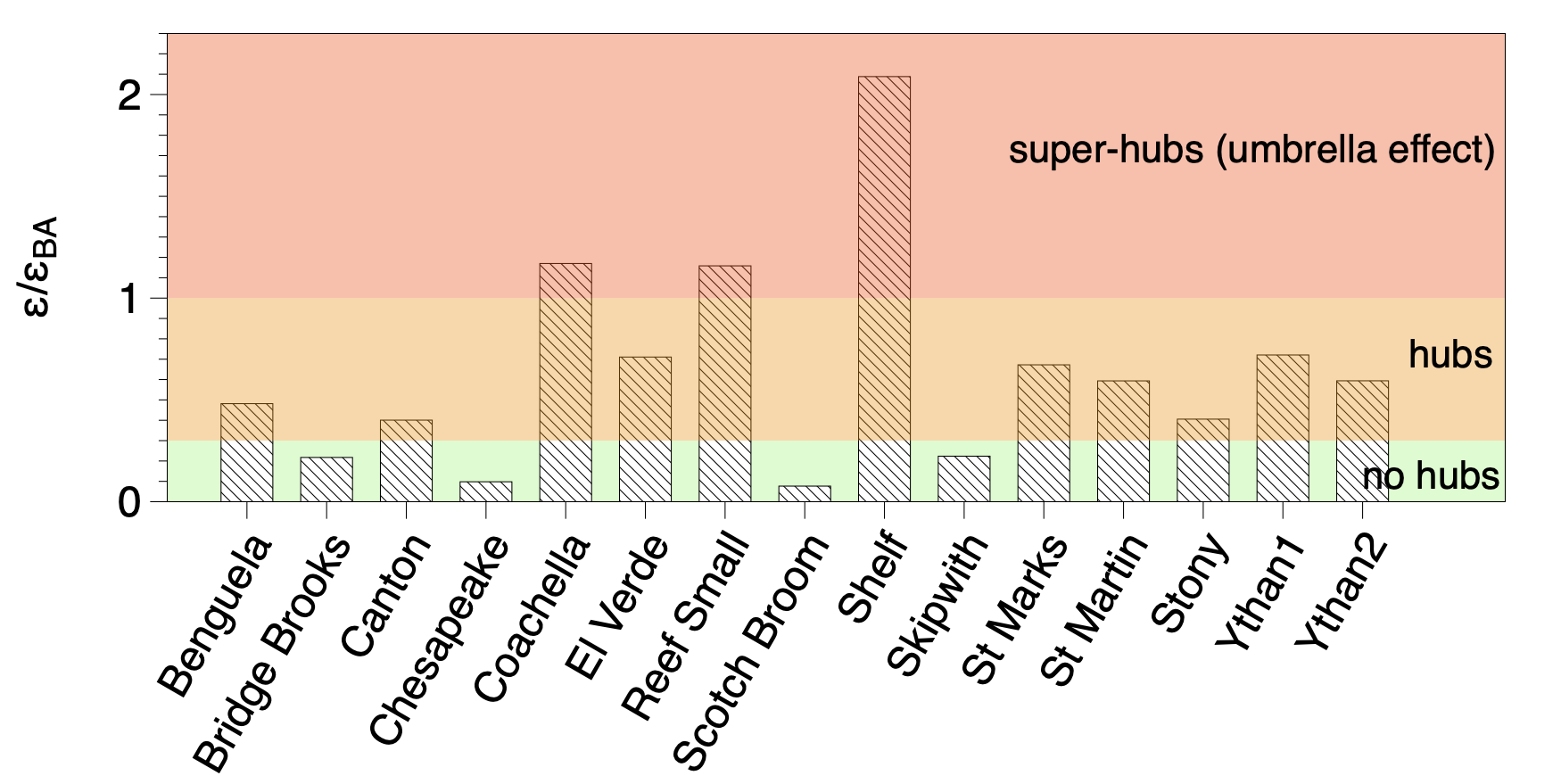}
\par\end{centering}
\caption{$\epsilon/\epsilon_{BA}$ for all 15 food webs. We have highlighted three regimes: $\epsilon/\epsilon_{BA}<0.5$, where the network is essentially not bypassing hubs,  $0.5<\epsilon/\epsilon_{BA}<1$, where the real network and its analogous BA network have fairly similar associated navigability gain and thus the real network is bypassing hubs, and $\epsilon/\epsilon_{BA}>1$ where the real network is bypassing hubs stronger than in the BA model. This last situation takes place when super-hubs emerge, i.e. nodes that connect to a large portion of the rest of nodes in the network.}
\label{fig:food}
\end{figure}

\begin{figure}[t!]
\begin{centering}
\includegraphics[width=0.7\textwidth]{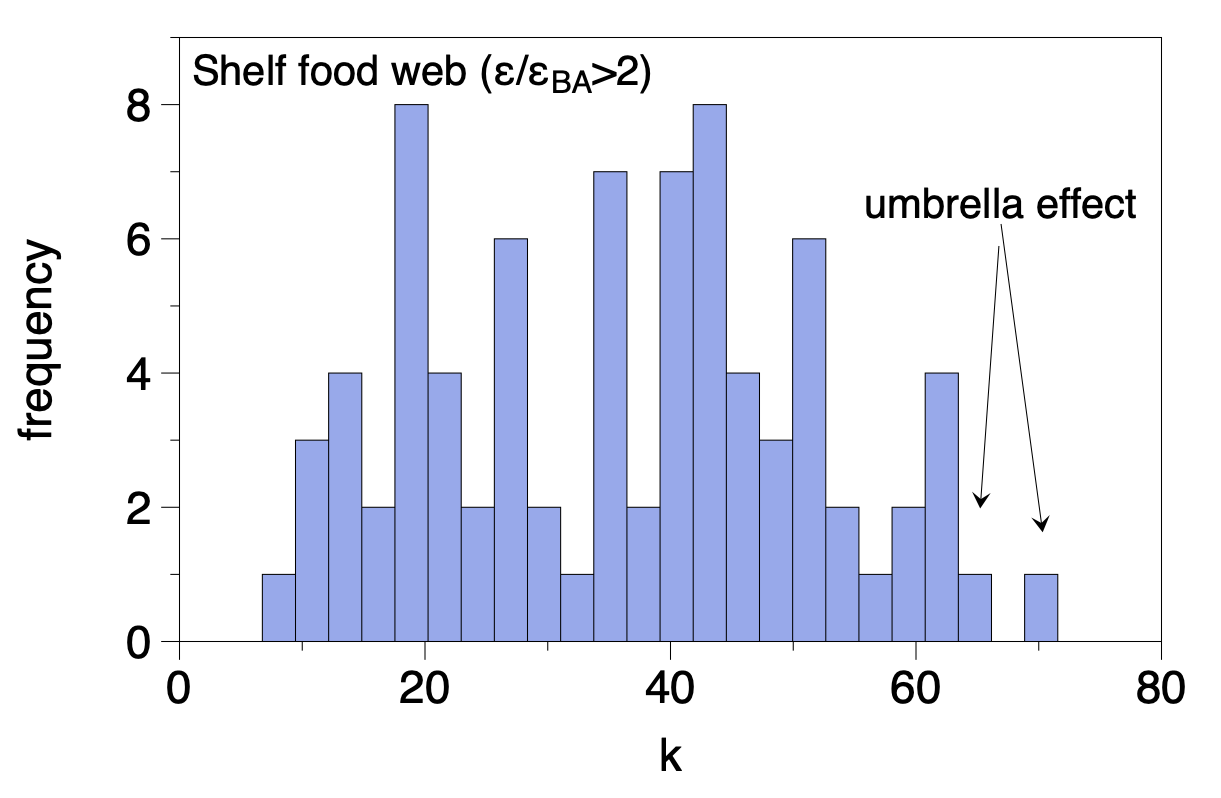}
\par\end{centering}
\caption{Degree frequency histogram of the Shelf food web (the one with $\epsilon/\epsilon_{BA}>2$). This network has $N=81$ nodes and a large number of them have very high degree. Some of these nodes will convert into super-hubs and then be likely to evidence the umbrella effect (see SI section S6) which make the navigability gain $\epsilon$ larger than the expected for a BA network.}
\label{fig:Shelf}
\end{figure}

\paragraph{S11.3 $\epsilon$ and $\epsilon/\epsilon_{BA}$: further analysis}
Here we give a closer inspection into the values of table 1 of the main article. Overall, we can see that  for a groups of networks (not necessarily domain specific), the bypass navigability gain $\epsilon$ is similar to the one expected for a BA model with the same number of nodes $N$ and mean degree $\langle k \rangle$, suggesting that the type of heterogeneity is similar and that the contribution of bypasses is substantial due to the presence of hubs. This is the case, for instance, in the Human brain, collaboration networks, C. Elegans neurons, or USA airports: for all these cases $\epsilon/\epsilon_{BA}>0.75$. \\
There is a second group of networks where such ratio is lower, suggesting that either the role of hubs is, at least, less prominent than what we would expect in a BA model with the same characteristics, or that even if hubs. exist, the network is designed to not bypasses. In this group we find a whole spectrum of values $\epsilon$: networks where the role of bypasses is still substantial but considerably less than in a comparable BA model (e.g. protein-protein interaction, with $\epsilon\approx 25$, with $\epsilon/\epsilon_{BA}\approx 0.55$), and networks where the bypasses have only a
marginal effect in the network's navigability (e.g. termite mounds, with $\epsilon\approx 3$, with $\epsilon/\epsilon_{BA}\approx 0.12$, or the power grid, with $\epsilon\approx 2.6$, with $\epsilon/\epsilon_{BA}\approx 0.05$). The result on the power grid is interesting: originally seen as an example of a scale-free network \cite{BA} (albeit with an exponent $4$, i.e. much more homogeneous than a BA network), its scale-freeness has been a matter of debate \cite{PG}. Nevertheless, it is clear that this network do have hubs. So our measures --that highlights that this network has bypasses have only about 5\% of the impact than a respective BA network would have-- determine that this network is not engineered to have routes bypassing hubs. This is in agreement to previous evidence on blackouts and cascading failures in such type of networks: these are examples of networks with fat-tailed degree distributions that do not follow an evolutionary process, which would certify the presence of bypasses.\\

\begin{figure}[t!]
\begin{centering}
\includegraphics[width=0.75\textwidth]{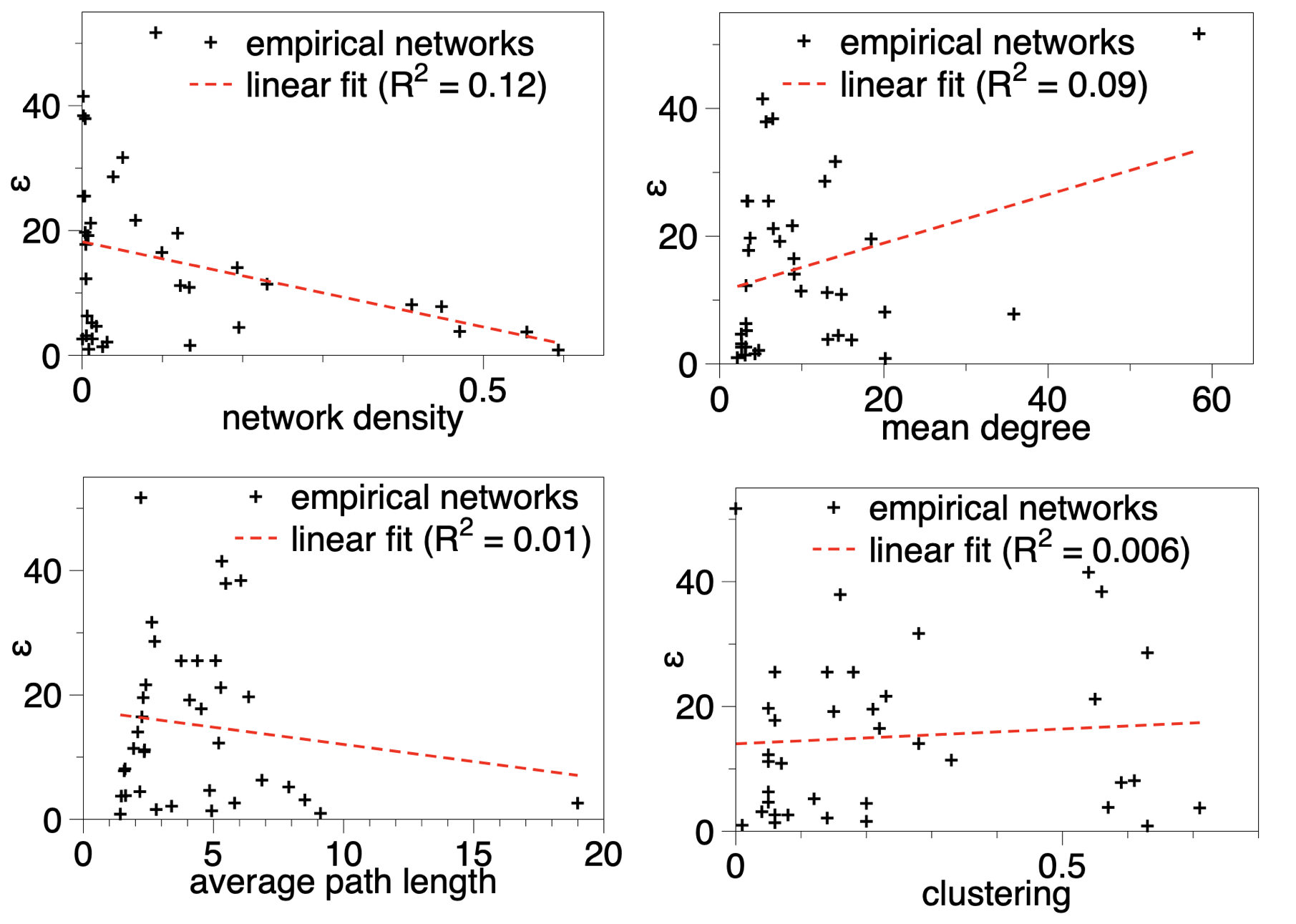}
\par\end{centering}
\caption{{Scatter plots of $\epsilon$ vs standard network metrics, along with linear regression estimates. In all cases the $R^2$ coefficient is rather low, certifying that $\epsilon$ is not trivially reduceable to more basic network measures.}}
\label{fig:correlations}
\end{figure}

\noindent Next, we disaggregate the calculation of $\epsilon/\epsilon_{BA}$ performed on the 15 food webs. In Fig. \ref{fig:food} we plot the estimate of $\epsilon/\epsilon_{BA}$ in each of them, where $\epsilon_{BA}$ is computed by generating BA models with the same number of nodes and mean degree that the empirical network, and averaging the resulting value of the navigability gain over 100 realizations of the BA model. We find food-webs across three regimes: the regime of low $\epsilon/\epsilon_{BA}$, which indicates that the network does not have hubs or  that the network has not been designed to bypass these, the regime where $\epsilon \approx \epsilon_{BA}$, which indicates that the network is compatible with a BA model and the navigability gain is therefore associated to hubs being bypassed, and a (new) third regime $\epsilon/\epsilon_{BA}$, found for very dense networks, where there are some nodes in the empirical network that likely evidence the umbrella effect. Figure \ref{fig:Shelf} illustrates this fact for the case of the Shelf food web ($\epsilon/\epsilon_{BA}>2$). In this figure we plot the degree frequency of this food web, showing that this is a notably dense network where we find some nodes with extreme degree (note that this network has $N=81$ nodes) which are candidates for the umbrella effect.

\paragraph{{S11.4 Lack of correlation between $\epsilon$ and standard network metrics}}
{In Fig.\ref{fig:correlations} we plot scatter plots of $\epsilon$ vs standard network metrics (network density, mean degree, average path length and average clustering) for all empirical networks analysed, showing that no trivial correlation emerges, specially in the region of lower density. In this sense, our results are not trivially reduceable to other standard network metrics.}

\bibliographystyle{naturemag}

\end{document}